\documentclass[conference]{IEEEtran}

\usepackage[utf8]{inputenc}
\usepackage{amsmath}
\usepackage{amssymb}
\usepackage{graphicx}
\usepackage{subcaption}
\usepackage{hyperref}
\usepackage{cite}
\usepackage{svg}
\usepackage{tikz}
\usepackage[obeyFinal]{todonotes}
\usepackage{xcolor}
\usepackage{booktabs}
\usetikzlibrary{arrows.meta,calc}

\begin{document}

\nocite{IEEEexample:BSTcontrol}

\title{
From sLLG to Fokker–Planck: Accurate WER Modeling for Non-Axisymmetric MRAM Devices
}
\author{
\IEEEauthorblockN{
F. Garcia Redondo\IEEEauthorrefmark{1}\IEEEauthorrefmark{5}\IEEEauthorrefmark{6},
T. Bhowmik\IEEEauthorrefmark{2}\IEEEauthorrefmark{4},
M. Gama Monteiro\IEEEauthorrefmark{2},
Y. Xiang\IEEEauthorrefmark{2},
J. Van Houdt\IEEEauthorrefmark{2}\IEEEauthorrefmark{3},  K. Temst\IEEEauthorrefmark{2}\IEEEauthorrefmark{4}, 
S. Rao\IEEEauthorrefmark{2}
}
\IEEEauthorblockA{\IEEEauthorrefmark{1}imec Cambridge, UK; \IEEEauthorrefmark{2}imec Leuven, BE; \IEEEauthorrefmark{3}Semiconductor Physics, KU Leuven; \IEEEauthorrefmark{4}Quantum Solid State Physics, KU Leuven.}
\IEEEauthorblockA{\IEEEauthorrefmark{5}F. Garcia Redondo and T. Bhowmik contributed equally to this work. \IEEEauthorrefmark{6}fernando.garciaredondo@imec-int.com}
}

\maketitle

\begin{abstract}
The Fokker--Planck (FP) equation is essential for predicting write error rates (WER) in STT and SOT-MRAM devices, but traditional 1D projections fail when symmetry is broken by in-plane fields, field-like torques, or anisotropic barriers. We develop a 2D finite-volume (FVM) solver on the unit sphere and validate it against $10^6$-trajectory stochastic Landau--Lifshitz--Gilbert (sLLG) simulations. The solver supports four discretization schemes---central, Scharfetter--Gummel (SG), upwind, and hybrid adaptive blending---each with different Péclet-dependent accuracy and monotonicity properties. We demonstrate that central differencing recovers ground-truth WER for STT and SOT geometries where 2D effects dominate, and show that the choice of discretization scheme directly affects predicted WER. For magnetic simulations, we recommend hybrid adaptive blending as the optimal balance of accuracy and stability across variable Péclet regimes. These results establish that customizable discretization is critical for accurate, unbiased predictions of switching dynamics in next-generation magnetic memory.
\end{abstract}


\section{Introduction}

Write error rate (WER) is the defining reliability metric for memory device technology validation, including MRAM. Predicting WER---especially below $10^{-6}$---via direct measurement is impractical; computational modeling is essential~\cite{Xie2017, garcia2021compact,garcia2021fokker, liu_high-accuracy_2025}. Accurate WER prediction is critical for circuit design and product development, enabling designers to assess memory reliability, optimize write margins, and predict failure rates across process and temperature variations. Beyond baseline reliability metrics, WER models are essential for analyzing complex physical phenomena such as magnetic immunity---the degradation of write reliability under external magnetic field perturbations---which directly impacts device robustness in real-world applications~\cite{meeren_magnetic_2024}.

\begin{figure}[!t]
\centering
\includegraphics[width=1\columnwidth]{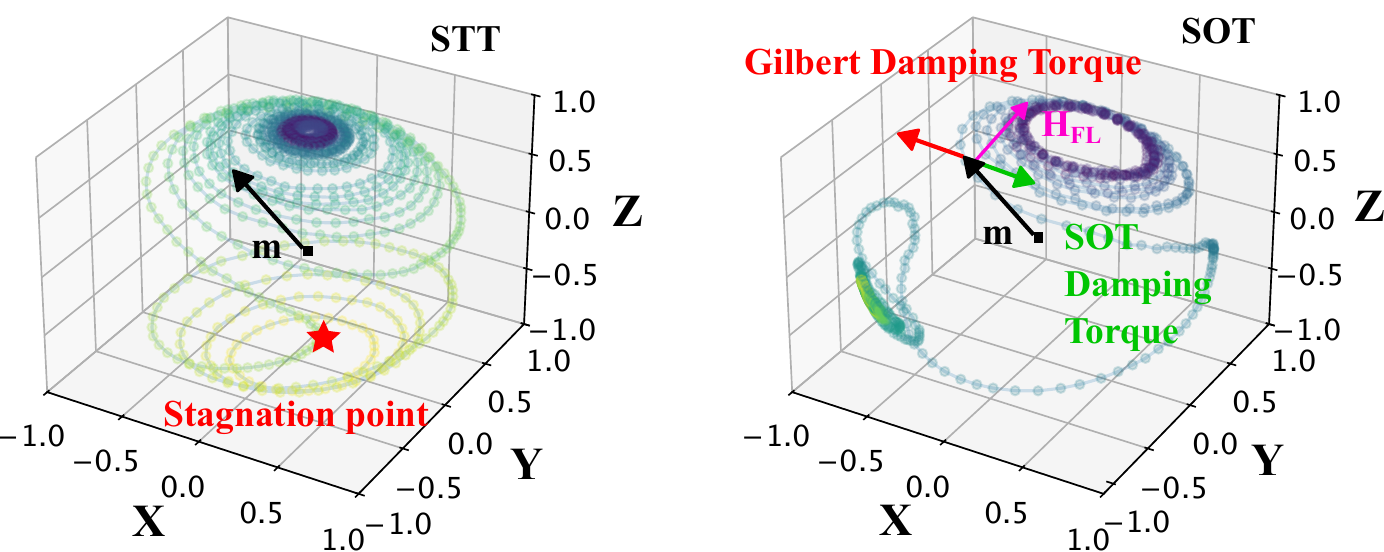}
\caption{STT/SOT MRAM's free layer magnetizations $\mathbf{m}$ on the unit sphere, driven by spin torque and thermal fluctuations. The azimuthally asymmetric in-plane field requires 2D treatment.
}
\label{fig:intro_mram}
\end{figure}

The standard approach collapses the Landau--Lifshitz--Gilbert--Slonczewski (LLGS) stochastic dynamics onto a single coordinate $z = \cos\theta$ and solves the 1D Fokker--Planck equation (FPE)~\cite{garcia2021fokker, liu_high-accuracy_2025, Butler2012, Torunbalci2018a}. This reduction is exact only under axial symmetry (energy and torques depend on $\theta$ alone).
This assumption breaks down for external in-plane fields ($h_x \neq 0$) which break azimuthal symmetry, field-like torques introducing $\phi$-dependent precession~\cite{song_spin-orbit_2021, bhowmik_modeling_2025}, SOT geometry where damping-like and field-like components explicitly couple $\theta$ and $\phi$, and asymmetric barriers: tilted easy axes or biaxial anisotropy --Figure~\ref{fig:intro_mram}.

As the computational time increase more than linearly \cite{garcia2021fokker} with the number of s-LLGS trajectories, obtaining product grade WER analyses becomes impractical. In all these cases the FPE 1D approximation systematically misrepresents switching dynamics a 2D treatment in ($\phi$, $\theta$) is required~\cite{Xie2017}.  However, in \cite{Xie2017} the authors do not discuss the suitability of finite-element methods (FEM) for MRAM probability distributions solving.
Similarly, the effects of boundary conditions and discretization schemes have so far never been investigated in the context of MRAM devices.
Moreover, compared with standard continuous finite-element discretizations, finite-volume methods (FVM) are particularly attractive for Fokker–Planck equations because they are naturally formulated in conservative flux form and can be designed to preserve the non-negativity of the probability density, two key properties for probability-conservation problems \cite{a_carrillo_finite-volume_2015}.

This work presents a 2D finite-volume solver validated against stochastic LLG for STT and SOT devices, and demonstrates how the choice of discretization scheme directly affects WER. Spatial discretization is not merely a numerical detail. Different flux schemes (e.g., central, Scharfetter--Gummel, and upwind) introduce different levels of numerical diffusion and exhibit varying accuracy across Péclet regimes. In 2D Fokker--Planck simulations, these differences can significantly bias predicted switching probabilities and WER. This work systematically compares commonly used discretization schemes for MRAM applications and demonstrates that appropriate scheme selection is essential for accurate reliability prediction.

\section{Methodology and Framework}

\subsection{s-LLGS Governing Equations and Solver Implementation}

The stochastic Landau--Lifshitz--Gilbert--Slonczewski (s-LLGS) equation is used as the trajectory-level reference model for thermally activated switching dynamics.Under the sign convention adopted here, a positive field-like torque coefficient ($\eta>0$) causes the field-like torque (FLT) to assist the damping-like torque (DLT) in overcoming Gilbert damping, thereby reducing the threshold switching current~\cite{song_spin-orbit_2021, bhowmik_modeling_2025, sispad_modeling_2026}. The magnetization unit vector $\mathbf{m}$ evolves according to \cite{donahue_object-oriented_2015} (I.S.U.):

\begin{equation}
\begin{aligned}
\frac{d\mathbf{m}}{dt}=\;&
-\gamma'\,\mathbf{m}\times\mathbf{H}_{\rm eff}
+\alpha\,\mathbf{m}\times\frac{d\mathbf{m}}{dt} \\
&-\gamma' \beta \varepsilon \,\mathbf{m}\times(\mathbf{m}\times\boldsymbol{\sigma})
+\gamma' \beta \varepsilon'\mathbf{m}\times\boldsymbol{\sigma},
\end{aligned}
\end{equation}

where $\gamma$ is the gyromagnetic ratio, $\mu_0$ is the vacuum permeability, $\alpha$ is the Gilbert damping constant, $\boldsymbol{\sigma}$ is the spin-polarization direction and $\gamma'=\frac{\gamma \mu_0}{1+\alpha^2}$. The effective field $\mathbf{H}_{\rm eff}$ includes perpendicular magnetic anisotropy, demagnetization, and any external field.
The spin torques for STT devices are
\begin{equation}
\varepsilon=
\frac{p_{STT}\Lambda^2}{(\Lambda^2+1)+(\Lambda^2-1)(\mathbf{m} \cdot \mathbf{\sigma})},
\beta=|\frac{\hbar}{\mu_0 e}| \frac{I_{STT}}{A_{FM}t_{FM} M_s},
\end{equation}

and for SOT devices
\begin{equation}
\begin{split}
    \varepsilon=
\frac{p_{SOT}\Lambda^2}{(\Lambda^2+1)+(\Lambda^2-1)(\mathbf{m} \cdot \mathbf{\sigma})},\\
\beta=|\frac{\hbar}{\mu_0 e}| \frac{I_{SOT}}{A_{SOT}t_{FM} M_s} ,\varepsilon'_{SOT}=\eta\varepsilon.
\end{split}
\end{equation}   
Here, $M_s$ is the saturation magnetization, $\Lambda$ is the spin-torque asymmetry parameter, $p_{SOT}/p_{STT}$ are the spin Hall angle and STT polarization factor, $I$ is the applied current, $t_{\rm FM}$ is the free-layer thickness and $A_{SOT}$ is the area of heavy metal track, and $A_{FM}$ refers to the area of magnetic free layer.

At finite temperature, thermal fluctuations are incorporated through a stochastic thermal field 
$\mathbf{H}_{\rm th}(t)$ added to $\mathbf{H}_{\rm eff}$:

\begin{equation}
\mathbf{H}_{\rm eff}
=
\mathbf{H}_{\rm det}
+
\mathbf{H}_{\rm th}(t),
\end{equation}
where $\mathbf{H}_{\rm det}$ contains anisotropy, demagnetization, and external fields. 
The thermal field $\mathbf{H}_{\rm th}$ is modeled as a Gaussian white noise~ \cite{garcia2021fokker, Butler2012, Torunbalci2018a} satisfying

\begin{equation}
\langle H_{{\rm th},i}(t)\rangle =0,
\end{equation}

\begin{equation}
\langle H_{{\rm th},i}(t)H_{{\rm th},j}(t')\rangle
=
\frac{2\alpha k_B T}
{\gamma \mu_0 M_s V}
\delta_{ij}\delta(t-t'),
\end{equation}
consistent with the fluctuation--dissipation theorem.

Thermal fluctuations are introduced through the standard stochastic effective field. The stochastic differential equation is interpreted and integrated in the \textit{Stratonovich} sense, consistent with multiplicative noise on the unit sphere~\cite{garcia2021compact}.

A parallel CPU/GPU implementation was developed using Google's \texttt{torchsde} package~\cite{kidger2021neuralsde}, following the open Python templates of~\cite{garcia2021compact,garcia2021fokker}. This solver provides the stochastic baseline used for all Fokker--Planck validation.


\subsection{Fokker--Planck Governing Equations}

The probability density $\rho(\theta,\phi,t)$ is defined per unit solid angle on the unit sphere and satisfies the normalization condition

\begin{equation}
\int_0^{2\pi}\int_0^\pi
\rho(\theta,\phi,t)\sin\theta\,d\theta\,d\phi=1.
\end{equation}

Under the corresponding Stratonovich interpretation, the stochastic
magnetization dynamics induce a probability density evolution on the unit sphere.
Applying standard stochastic Fokker--Planck theory yields to the corresponding probability transport equation, which is written in conservative form as

\begin{equation}
\frac{\partial \rho}{\partial t}
+\nabla_{\rm S}\cdot\mathbf{J}=0,
\end{equation}

where $\nabla_{\rm S}$ denotes the surface divergence operator and the probability flux is

\begin{equation}
\mathbf{J}=
\mathbf{v}\rho-D\nabla_{\rm S}\rho.
\label{eq:J}
\end{equation}

Here $\mathbf{v}(\theta,\phi)$ is the deterministic drift velocity induced by the LLGS torque terms, and $D$ is the isotropic thermal diffusion coefficient given by the fluctuation--dissipation relation \cite{Butler2012, Xie2017}:

\begin{equation}
D=
\frac{\alpha\gamma k_B T}
{(1+\alpha^2)\mu_0 M_s V},
\end{equation}

where $V$ is the free-layer volume.

The FP equation is the density-level representation corresponding to the Stratonovich s-LLGS dynamics described above.


\subsection{Finite Volume Discretization on the Sphere}

\begin{figure}[!h]
\centering
\def\R{5}              
\def\FigScale{0.6}     
\def\angEl{22}         

\pgfmathsetmacro{\MainFont}{20*\FigScale}
\pgfmathsetmacro{\MainFontLead}{1.2*\MainFont}
\pgfmathsetmacro{\SmallFont}{20*\FigScale}
\pgfmathsetmacro{\SmallFontLead}{1.2*\SmallFont}

\pgfmathsetmacro{\GridLW}{0.50*\FigScale}
\pgfmathsetmacro{\MainLW}{1.15*\FigScale}
\pgfmathsetmacro{\AxisLW}{1.10*\FigScale}
\pgfmathsetmacro{\OutlineLW}{0.90*\FigScale}

\pgfmathsetmacro{\ArrowLen}{4.2*\FigScale}
\pgfmathsetmacro{\ArrowWid}{2.9*\FigScale}

\newcommand\pgfmathsinandcos[3]{%
  \pgfmathsetmacro#1{sin(#3)}%
  \pgfmathsetmacro#2{cos(#3)}%
}

\newcommand\LongitudePlane[3][current plane]{%
  \pgfmathsinandcos\sinEl\cosEl{#2}%
  \pgfmathsinandcos\sint\cost{#3}%
  \tikzset{#1/.estyle={cm={\cost,\sint*\sinEl,0,\cosEl,(0,0)}}}%
}

\newcommand\LatitudePlane[3][current plane]{%
  \pgfmathsinandcos\sinEl\cosEl{#2}%
  \pgfmathsinandcos\sint\cost{#3}%
  \pgfmathsetmacro\yshift{\cosEl*\sint}%
  \tikzset{#1/.estyle={cm={\cost,0,0,\cost*\sinEl,(0,\yshift)}}}%
}

\newcommand\DrawLongitudeCircle[3][1]{%
  \LongitudePlane{\angEl}{#2}
  \tikzset{current plane/.prefix style={scale=#1}}
  \pgfmathsetmacro\angVis{atan(sin(#2)*cos(\angEl)/sin(\angEl))}
  \draw[current plane,#3]
    (\angVis:1) arc (\angVis:\angVis+180:1);
  \draw[current plane,#3,densely dashed,opacity=0.7]
    (\angVis-180:1) arc (\angVis-180:\angVis:1);
}

\newcommand\DrawLatitudeCircle[3][1]{%
  \LatitudePlane{\angEl}{#2}
  \tikzset{current plane/.prefix style={scale=#1}}
  \pgfmathsetmacro\sinVis{sin(#2)/cos(#2)*sin(\angEl)/cos(\angEl)}
  \pgfmathsetmacro\angVis{asin(min(1,max(\sinVis,-1)))}
  \draw[current plane,#3]
    (\angVis:1) arc (\angVis:-\angVis-180:1);
  \draw[current plane,#3,densely dashed,opacity=0.7]
    (180-\angVis:1) arc (180-\angVis:\angVis:1);
}

\begin{tikzpicture}[
    scale=\FigScale,
    transform shape,
    >=Stealth,
    line cap=round,
    line join=round,
    every node/.style={
      font=\fontsize{\MainFont pt}{\MainFontLead pt}\selectfont
    }
  ]

\def\lonA{-32}
\def\lonB{-14}
\def\lonC{0}
\def\lonD{14}
\def\lonE{32}

\def\latEq{-10}
\def\latNorth{20}
\def\latUpper{48}
\def\latSouth{-42}
\def\latLower{-62}

\definecolor{meridblue}{RGB}{70,125,255}
\definecolor{latgreen}{RGB}{35,165,55}
\definecolor{forceblue}{RGB}{25,95,230}
\definecolor{sphereedge}{RGB}{95,95,95}

\pgfmathsetmacro{\H}{\R*cos(\angEl)}

\pgfmathsetmacro{\cellxb}{0.136*\R}
\pgfmathsetmacro{\cellxt}{0.112*\R}
\pgfmathsetmacro{\cellyb}{-0.116*\R}
\pgfmathsetmacro{\cellyt}{ 0.108*\R}
\pgfmathsetmacro{\celldot}{-0.046*\R}

\pgfmathsetmacro{\Wx}{-0.436*\R}
\pgfmathsetmacro{\Wy}{ 0.010*\R}
\pgfmathsetmacro{\Ex}{ 0.436*\R}
\pgfmathsetmacro{\Ey}{ 0.010*\R}
\pgfmathsetmacro{\Nx}{ 0.0}
\pgfmathsetmacro{\Ny}{ 0.392*\R}
\pgfmathsetmacro{\Sx}{ 0.0}
\pgfmathsetmacro{\Sy}{-0.410*\R}

\pgfmathsetmacro{\nodeR}{0.054*\R}
\pgfmathsetmacro{\nodeDot}{0.011*\R}
\pgfmathsetmacro{\centerDot}{0.015*\R}

\pgfmathsetmacro{\xaxA}{-0.65*\R}
\pgfmathsetmacro{\xayA}{-0.35*\R}
\pgfmathsetmacro{\xaxB}{-0.90*\R}
\pgfmathsetmacro{\xayB}{-0.54*\R}

\pgfmathsetmacro{\thx}{1.08*\R}
\pgfmathsetmacro{\thy}{0.06*\R}
\pgfmathsetmacro{\thrx}{0.188*\R}
\pgfmathsetmacro{\thry}{0.236*\R}

\pgfmathsetmacro{\phix}{0.61*\R}
\pgfmathsetmacro{\phiy}{-0.694*\R}
\pgfmathsetmacro{\phirx}{0.256*\R}
\pgfmathsetmacro{\phiry}{0.164*\R}

\pgfmathsetmacro{\WLx}{-0.66*\R}
\pgfmathsetmacro{\WLy}{ 0.07*\R}
\pgfmathsetmacro{\WLix}{-0.66*\R}
\pgfmathsetmacro{\WLiy}{-0.01*\R}

\pgfmathsetmacro{\ELx}{0.64*\R}
\pgfmathsetmacro{\ELy}{0.07*\R}
\pgfmathsetmacro{\ELix}{0.66*\R}
\pgfmathsetmacro{\ELiy}{-0.01*\R}

\pgfmathsetmacro{\NLy}{0.60*\R}
\pgfmathsetmacro{\NLiy}{0.504*\R}
\pgfmathsetmacro{\SLy}{-0.584*\R}
\pgfmathsetmacro{\SLiy}{-0.70*\R}

\pgfmathsetmacro{\FWax}{-0.364*\R}
\pgfmathsetmacro{\FWay}{ 0.004*\R}
\pgfmathsetmacro{\FWtx}{-0.228*\R}
\pgfmathsetmacro{\FWty}{ 0.032*\R}

\pgfmathsetmacro{\FEax}{ 0.364*\R}
\pgfmathsetmacro{\FEay}{ 0.004*\R}
\pgfmathsetmacro{\FEtx}{ 0.228*\R}
\pgfmathsetmacro{\FEty}{ 0.032*\R}

\pgfmathsetmacro{\FNay}{0.336*\R}
\pgfmathsetmacro{\FNtx}{0.034*\R}
\pgfmathsetmacro{\FNty}{0.236*\R}

\pgfmathsetmacro{\FSay}{-0.36*\R}
\pgfmathsetmacro{\FStx}{0.034*\R}
\pgfmathsetmacro{\FSty}{-0.252*\R}

\pgfmathsetmacro{\thlabx}{1.14*\R}
\pgfmathsetmacro{\thlaby}{0.14*\R}
\pgfmathsetmacro{\philabx}{0.86*\R}
\pgfmathsetmacro{\philaby}{-0.596*\R}

\shade[ball color=gray!10,opacity=0.45] (0,0) circle (\R);
\draw[sphereedge,line width=\OutlineLW pt] (0,0) circle (\R);

\coordinate (O) at (0,0);
\coordinate (Npole) at (0,\H);
\coordinate (Spole) at (0,-\H);

\DrawLongitudeCircle[\R]{\lonA}{latgreen,line width=\GridLW pt}
\DrawLongitudeCircle[\R]{\lonB}{latgreen,line width=\GridLW pt}
\DrawLongitudeCircle[\R]{\lonC}{latgreen,line width=\GridLW pt}
\DrawLongitudeCircle[\R]{\lonD}{latgreen,line width=\GridLW pt}
\DrawLongitudeCircle[\R]{\lonE}{latgreen,line width=\GridLW pt}

\DrawLatitudeCircle[\R]{\latEq}{meridblue,line width=\GridLW pt}
\DrawLatitudeCircle[\R]{\latNorth}{meridblue,line width=\GridLW pt}
\DrawLatitudeCircle[\R]{\latUpper}{meridblue,line width=\GridLW pt}
\DrawLatitudeCircle[\R]{\latSouth}{meridblue,line width=\GridLW pt}
\DrawLatitudeCircle[\R]{\latLower}{meridblue,line width=\GridLW pt}

\draw[-{Stealth[length=\ArrowLen pt,width=\ArrowWid pt]},black,line width=\AxisLW pt]
  (0,\H) -- (0,\H+0.18*\R) node[above] {$z$};

\draw[-{Stealth[length=\ArrowLen pt,width=\ArrowWid pt]},black,line width=\AxisLW pt]
  (\R,0) -- (\R+0.20*\R,0) node[right] {$y$};

\draw[-{Stealth[length=\ArrowLen pt,width=\ArrowWid pt]},black,line width=\AxisLW pt]
  (\xaxA,\xayA) -- (\xaxB,\xayB) node[left] {$x$};

\path[fill=blue!10,draw=black!70,line width=\MainLW pt]
  (-\cellxb,\cellyb) --
  ( \cellxb,\cellyb) --
  ( \cellxt,\cellyt) --
  (-\cellxt,\cellyt) -- cycle;

\fill (0,\celldot) circle (\centerDot);
\node at (0,0.006*\R) {$(i,j)$};

\coordinate (W)  at (\Wx,\Wy);
\coordinate (E)  at (\Ex,\Ey);
\coordinate (NN) at (\Nx,\Ny);
\coordinate (SS) at (\Sx,\Sy);

\draw[forceblue,line width=\MainLW pt] (W) circle (\nodeR);
\fill[forceblue] (W) circle (\nodeDot);
\node at (\WLx,\WLy) {\textbf{W}};
\node[font=\fontsize{\SmallFont pt}{\SmallFontLead pt}\selectfont] at (\WLix,\WLiy) {$(i\!-\!1,j)$};

\draw[forceblue,line width=\MainLW pt] (E) circle (\nodeR);
\fill[forceblue] (E) circle (\nodeDot);
\node at (\ELx,\ELy) {\textbf{E}};
\node[font=\fontsize{\SmallFont pt}{\SmallFontLead pt}\selectfont] at (\ELix,\ELiy) {$(i\!+\!1,j)$};

\draw[latgreen,line width=\MainLW pt] (NN) circle (\nodeR);
\fill[latgreen] (NN) circle (\nodeDot);
\node at (0,\NLy) {\textbf{N}};
\node[font=\fontsize{\SmallFont pt}{\SmallFontLead pt}\selectfont] at (0,\NLiy) {$(i,j\!+\!1)$};

\draw[latgreen,line width=\MainLW pt] (SS) circle (\nodeR);
\fill[latgreen] (SS) circle (\nodeDot);
\node at (0,\SLy) {\textbf{S}};
\node[font=\fontsize{\SmallFont pt}{\SmallFontLead pt}\selectfont] at (0,\SLiy) {$(i,j\!-\!1)$};

\draw[-{Stealth[length=\ArrowLen pt,width=\ArrowWid pt]},forceblue,line width=\MainLW pt]
  (\FWax,\FWay) -- (-\cellxb,0);
\node[above] at (\FWtx,\FWty) {$F_W$};

\draw[-{Stealth[length=\ArrowLen pt,width=\ArrowWid pt]},forceblue,line width=\MainLW pt]
  (\cellxb,0) -- (\FEax,\FEay);
\node[above] at (\FEtx,\FEty) {$F_E$};

\draw[-{Stealth[length=\ArrowLen pt,width=\ArrowWid pt]},latgreen,line width=\MainLW pt]
  (0,\cellyt) -- (0,\FNay);
\node[right] at (\FNtx,\FNty) {$F_N$};

\draw[-{Stealth[length=\ArrowLen pt,width=\ArrowWid pt]},latgreen,line width=\MainLW pt]
  (0,\cellyb) -- (0,\FSay);
\node[right] at (\FStx,\FSty) {$F_S$};

\draw[-{Stealth[length=\ArrowLen pt,width=\ArrowWid pt]},latgreen,line width=\MainLW pt]
  (\thx,\thy) arc[start angle=0,end angle=62,x radius=\thrx,y radius=\thry];
\node[latgreen] at (\thlabx,\thlaby) {$\theta$};

\draw[-{Stealth[length=\ArrowLen pt,width=\ArrowWid pt]},forceblue,line width=\MainLW pt]
  (\phix,\phiy) arc[start angle=-112,end angle=-34,x radius=\phirx,y radius=\phiry];
\node[forceblue] at (\philabx,\philaby) {$\phi$};

\end{tikzpicture}

\caption{Finite-volume stencil on the unit sphere. Control volume $(i,j)$ exchanges numerical flux with its four nearest neighbors in the structured $(\phi,\theta)$ discretization. Face lengths and cell areas incorporate the spherical metric directly through the mesh geometry. All four discretization schemes---Central, Scharfetter--Gummel, Upwind, and Hybrid---differ only in the approximation of these face fluxes.}
\label{fig:methods_stencil}
\end{figure}
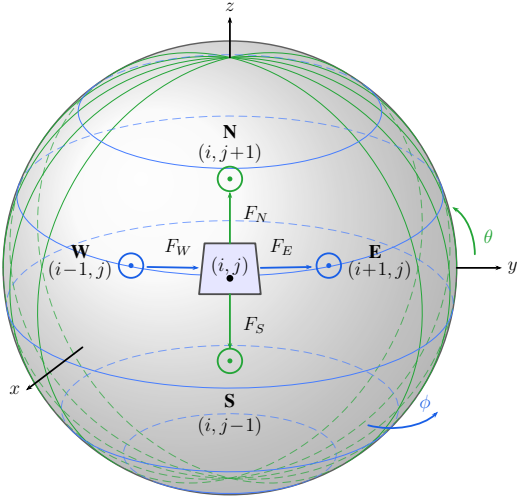

The FP equation is discretized on a structured $(\phi,\theta)$ mesh with $N_\phi\times N_\theta$ control volumes --Figure~\ref{fig:methods_stencil}. Integrating over each cell $\Omega_k$ gives

\begin{equation}
\frac{d}{dt}\int_{\Omega_k}\rho\,dA
=
-\oint_{\partial\Omega_k}\mathbf{J}\cdot\hat{n}\,dl,
\end{equation}

which is approximated by the sum of numerical face fluxes.

This finite-volume formulation is conservative by construction: total probability is preserved to solver tolerance for closed domains, with probability loss occurring only through absorbing boundaries.

The spherical metric is handled geometrically through exact face lengths and cell areas, rather than through explicit pointwise division by $\sin\theta$. This avoids coordinate singularities near the poles and improves numerical robustness.

For a cell indexed by $(i,j)$:

\begin{itemize}
\item \textbf{Azimuthal faces} (constant $\phi$, coupling neighboring $\theta$ cells):
\begin{equation}
A_\phi=\Delta\theta_j.
\end{equation}

\item \textbf{Polar faces} (constant $\theta$, coupling neighboring $\phi$ cells):
\begin{equation}
A_\theta=\Delta\phi_i\sin\theta_{\rm face}.
\end{equation}

\item \textbf{Cell area}:
\begin{equation}
A_{ij}=
\Delta\phi_i
\left(
\cos\theta_{j-\frac12}
-
\cos\theta_{j+\frac12}
\right).
\end{equation}
\end{itemize}

These expressions exactly represent the surface measure on the unit sphere. Drift and diffusion coefficients are supplied without embedded metric factors; all geometry is carried by the mesh itself.


\subsection{Boundary Conditions and Write Error Rate}

Boundary conditions determine how probability flux interacts with the computational domain and are essential for physical fidelity, conservation, and numerical robustness. The solver supports four boundary types on each coordinate direction:

\begin{enumerate}
\item \textbf{Reflecting}: zero normal probability flux,
\item \textbf{Absorbing}: outward probability is removed,
\item \textbf{Periodic}: opposite boundaries are directly identified,
\item \textbf{Antipodal}: opposite boundaries are identified with a half-period phase shift.
\end{enumerate}

Boundary conditions may be prescribed independently on each side of the $(\phi,\theta)$ domain.

\subsubsection{Boundary Conditions Used in This Work}

For spherical magnetization dynamics, the azimuthal angle is cyclic. Accordingly, all simulations reported here use periodic boundaries in $\phi$:

\begin{equation}
\rho(\theta,0,t)=\rho(\theta,2\pi,t).
\end{equation}

In the polar direction ($\theta$), two physically relevant closures are considered:

\begin{itemize}
\item \textbf{Reflecting in $\theta$:} zero normal flux at $\theta=0$ and $\theta=\pi$, producing a closed conservative domain.

\item \textbf{Antipodal in $\theta$:} pole crossings are identified through the spherical topology. Probability exiting through a pole re-enters with azimuth shifted by $\pi$:
\begin{align}
(\phi,0) &\equiv (\phi+\pi,0),\\
(\phi,\pi) &\equiv (\phi+\pi,\pi).
\end{align}
This avoids treating the poles as artificial walls and preserves smooth transport across coordinate singularities.
\end{itemize}

Unless otherwise stated, production runs use periodic boundaries in $\phi$ together with either reflecting or antipodal closure in $\theta$, depending on the experiment.

\subsubsection{Finite-Volume Boundary Implementation}

All boundaries are imposed through the same numerical face-flux operator used in the interior, ensuring consistency with the finite-volume formulation.

\begin{itemize}
\item \textbf{Reflecting:} face flux is set to zero.
\item \textbf{Periodic:} cells on opposite seams are directly coupled.
\item \textbf{Antipodal:} seam cells are coupled to azimuthal index $i+N_\phi/2 \pmod{N_\phi}$ (requiring even $N_\phi$).
\item \textbf{Absorbing:} outward flux is removed and accumulated separately.
\end{itemize}

Reflecting, periodic, and antipodal boundaries are conservative; total probability changes only through absorbing boundaries.

\subsubsection{Write Error Rate Definition}

For switching calculations, the switched state may be represented by an absorbing south-pole boundary or absorbing cap near $\theta=\pi$. The cumulative escaped probability is denoted $P_{\rm sw}(t)$, and the write error rate is

\begin{equation}
\mathrm{WER}(t)=1-P_{\rm sw}(t).
\end{equation}

Thus, the stochastic switching problem is reformulated as a first-passage probability calculation.


\subsection{Spatial Discretization Schemes}
For each control-volume face, the continuous flux in Eq.~(\ref{eq:J})
is projected onto the outward normal direction and discretized as a scalar
numerical flux $F=\mathbf{J}\cdot\hat{n}$.
The schemes below differ only in how this face flux is approximated.

At each face, the local transport regime is characterized by the Péclet number

\begin{equation}
\mathrm{Pe}(\theta,\phi)=
\frac{|\mathbf{v}|\Delta s}{D},
\label{eq:peclet}
\end{equation}

where $\Delta s$ is the face-normal distance. Low Péclet number indicates diffusion-dominated transport, while high Péclet number indicates drift-dominated transport.

Three numerical face-flux schemes are implemented.

\subsubsection{Central Difference}

\begin{equation}
F=
v\frac{\rho_L+\rho_R}{2}
-\frac{D}{\Delta s}(\rho_R-\rho_L).
\end{equation}

This scheme is second-order accurate on uniform meshes and minimally dissipative for smooth solutions, but may generate oscillatory error when transport is strongly under-resolved.

\subsubsection{Scharfetter--Gummel (SG)}

\begin{equation}
F=
\frac{D}{\Delta s}
\left[
B(-\mathrm{Pe})\rho_L
-
B(\mathrm{Pe})\rho_R
\right],
B(x)=\frac{x}{e^x-1}.
\end{equation}

SG is monotone over a wide Péclet range and widely used for drift--diffusion transport~\cite{scharfetter1969,selberherr1984}.

\subsubsection{Upwind}

\begin{equation}
F=
v\rho_{\rm upwind}
-\frac{D}{\Delta s}(\rho_R-\rho_L).
\end{equation}

Upwind is the most robust but also the most numerically dissipative of the three schemes.

\subsubsection{Hybrid (Smart Adaptive Blending)}
\label{sec:hybrid}
To balance accuracy across varying Péclet regimes, a hybrid blending strategy adaptively weights central and SG discretizations based on local Pe~\cite{jiang_hybrid_2016}:

\begin{equation}
F = (1-\theta) F_{\text{CD}} + \theta F_{\text{SG}},
\end{equation}

where the blending function $\theta(\mathrm{Pe})$ is defined via Hermite interpolation:

\begin{equation}
\theta(\mathrm{Pe}) = \begin{cases}
0 & \text{if } \mathrm{Pe} < 1.0 \\
3x^2 - 2x^3 & \text{if } 1.0 \le \mathrm{Pe} \le 2.0, \quad x = \mathrm{Pe} - 1 \\
0 & \text{if } \mathrm{Pe} > 2.0
\end{cases}
\end{equation}

This inverted blending (preferring central over wide Pe range) is motivated by the observation that in 2D spherical geometry with drift, the central scheme remains stable and accurate even at $\mathrm{Pe} \gg 1$. The hybrid formulation serves as a diagnostic: in our test cases (Section~\ref{sec:results_hybrid}), it naturally devolves ranges around 92.5-85.5\% central and 7.5-14.5\% blended regions, with SG never selected. This validates that central differencing is geometrically optimal for the coupled 2D transport.

\subsection{Temporal Integration}

Time advancement uses the $\theta$-method~\cite{strikwerda2004,morton2005}, which unifies explicit, implicit, and Crank--Nicolson schemes:

\begin{itemize}
\item $\theta=0$: explicit Euler,
\item $\theta=0.5$: Crank--Nicolson,
\item $\theta=1$: fully implicit Euler.
\end{itemize}

Crank--Nicolson is second-order accurate in time and unconditionally stable for linear diffusion operators, while fully implicit stepping is first-order accurate and strongly robust for stiff switching problems.


\subsection{Stability and CFL Constraints}
\label{sec:stability}

\subsubsection{Advection (Drift) Stability}

The advective Courant-Friedrichs-Lewy (CFL) condition characterizes the maximum transport distance per time step:

\begin{equation}
\mathrm{CFL}_{\rm adv} = \left| \mathbf{v}_\phi \right| \frac{\Delta t}{\Delta s_\phi} + \left| \mathbf{v}_\theta \right| \frac{\Delta t}{\Delta s_\theta},
\label{eq:cfl}
\end{equation}

where $\mathbf{v}_\phi$ and $\mathbf{v}_\theta$ are the azimuthal and polar drift velocities, and $\Delta s_\phi$, $\Delta s_\theta$ are the grid spacings.

Physically, $\mathrm{CFL}_{\rm adv}$ measures the number of grid cells traversed by drifting particles in a single time step. For \textbf{explicit} schemes ($\theta < 0.5$), unconditional stability requires $\mathrm{CFL}_{\rm adv} \leq 1$. For \textbf{implicit and Crank--Nicolson} ($\theta \geq 0.5$), no hard stability limit applies; however, large CFL values ($\gg 1$) can cause probability transport across many cells per step, leading to smearing and loss of accuracy. In practice, we maintain $\max(\mathrm{CFL}_{\rm adv}) \lesssim 0.5$ to preserve solution accuracy even in the implicit regime.

\subsubsection{Diffusion Stability}

Diffusive stability for explicit methods requires

\begin{equation}
\Delta t_{\text{diff,ex}} \leq \frac{1}{4D} \min(\Delta s_\phi^2, \Delta s_\theta^2).
\end{equation}

Implicit and Crank--Nicolson stepping remove this restriction entirely, permitting larger time steps without diffusive instability.

\subsubsection{Implications When Simulating Magnetic Memories}

For the magnetic switching problems investigated here --Section~\ref{sec:results}, we employ Crank--Nicolson ($\theta=0.5$) with a fixed time step $\Delta t = 10^{-3}$. This choice provides:
\begin{itemize}
  \item \textbf{Unconditional stability}: no CFL restriction,
  \item \textbf{Second-order accuracy in time}: $O(\Delta t^2)$ truncation error,
  \item \textbf{Robustness}: implicit treatment of stiff drift and diffusion operators,
  \item \textbf{Efficiency}: a single sparse linear solve per step, feasible on standard compute.
\end{itemize}

The solver automatically reports $\max(\mathrm{CFL}_{\rm adv})$ and local Péclet statistics for validation and mesh/timestep guidance. If large CFL values are observed (e.g., $>1$ over a significant fraction of the domain), refining $\Delta t$ or refining the mesh in high-drift regions is recommended to maintain solution accuracy and prevent feature smearing.


\subsection{Validation Methodology}

Ground truth is established using $N=10^6$ independent s-LLGS trajectories integrated with a stochastic midpoint method consistent with Stratonovich calculus~\cite{gardiner2009,kloeden1992}. The empirical write error rate is

\begin{equation}
\widehat{\mathrm{WER}}(t)=
1-\frac{1}{N}
\sum_{k=1}^{N}
\mathbf{1}[m_{z,k}(t)<0].
\end{equation}

The stochastic and FP solvers use identical physical parameters ($\alpha$, $\Delta_B$, torque efficiency, applied fields, and temperature). Therefore, any discrepancy in $\mathrm{WER}(t)$ arises from numerical approximation rather than model mismatch.

Validation is performed by comparing:

\begin{itemize}
\item $\mathrm{WER}(t)$ trajectories,
\item switching-time distributions,
\item intermediate density snapshots $\rho(\theta,\phi,t)$,
\item conservation and mesh-convergence behavior.
\end{itemize}

\section{Results and Discussion}
\label{sec:results}

\begin{table*}[t!]
\centering
\caption{MTJ characteristics and simulation parameters for STT and SOT devices. Formalism using \cite{donahue_object-oriented_2015, sispad_modeling_2026}.}
\begin{tabular}{@{}p{4.0cm}p{3.5cm}p{4.0cm}p{3.5cm}@{}}
\toprule
\textbf{STT Device} & & \textbf{SOT Device} & \\
Parameter & Scenarios A, B Value & Parameter & Scenario C Value \\
\midrule
$M_s$ [A/m] & $9.000 \times 10^{5}$ & $M_s$ [A/m] & $9.000 \times 10^{5}$ \\
$\alpha$ & $0.0086$ & $\alpha$ & $0.0086$ \\
$t_{\mathrm{fl}}$ [m] & $1.900 \times 10^{-9}$ & $t_{\mathrm{fl}}$ [m] & $1.200 \times 10^{-9}$ \\
$k_i$ [J/m$^2$] & $1.053 \times 10^{-3}$ & $k_i$ [J/m$^2$] & $6.648 \times 10^{-4}$ \\
$\Delta E$ [$k_B T$] & $58.98$ & $\Delta E$ [$k_B T$] & $17.02$ \\
$\tau_d$ [s] & $2.354 \times 10^{-9}$ & $\tau_d$ [s] & $2.484 \times 10^{-9}$ \\
$H_K^{\mathrm{eff}}$ [A/m] & $2.234 \times 10^{5}$ & $H_K^{\mathrm{eff}}$ [A/m] & $2.116 \times 10^{5}$ \\
$I_{c0}$ [A] & $2.128 \times 10^{-5}$ & $I_{c,\mathrm{SOT}}$ [A] & $0.00112$ \\
$\sigma_{\mathrm{STT}}$ & \{$0.0,0.0,1.0$\} & $\sigma_{\mathrm{SOT}}$ & \{$0.0,1.0,0.0$\} \\
$\Lambda_{\mathrm{STT}}$ & $1.2$ & $\Lambda_{\mathrm{SOT}}$ & $1.0$ \\
& & $\eta$ & $0.30$ \\
$p_{\mathrm{STT}}$ & $0.600$ & $p_{\mathrm{SOT}}$ & $0.30$ \\
$\varepsilon'_{\mathrm{STT}}$ & $0.003$ & $\varepsilon'_{\mathrm{SOT}}$ $(\eta\varepsilon_{\mathrm{SOT}})$ & $0.045$ \\
$I_{\mathrm{STT}}$ [A] & $-5.000 \times 10^{-5}$ & $I_{\mathrm{SOT}}$ [A] & $-1.680 \times 10^{-4}$ \\
& & $T_{DL}$ [$\mathrm{m}^{-1}$] & $1.889 \times 10^{8}$ \\
$I_{\mathrm{STT}}/I_{c0}$ & $-2.350$ & $h_s\;(\beta\varepsilon_{\mathrm{SOT}}/H_K^{\mathrm{eff}})$ & $-0.15$ \\
$\mathbf{H}_{\mathrm{ext}}$ [A/m] &
$(0,\,0,\,0),\,(2.69\times10^{4},\,0,\,0)$ &
$\mathbf{H}_{\mathrm{ext}}$ [A/m] &
$(3.17,\,3.17,\,3.17)\times10^{4}$ \\
$\mathbf{h}_{\mathrm{norm}}$ &
$(0,\,0,\,0),\,(0.15,\,0,\,0)$ &
$\mathbf{h}_{\mathrm{norm}}$ &
$(0.15,\,0.15,\,0.15)$ \\
\bottomrule
\end{tabular}
\label{tab:parameters}
\end{table*}

\subsection{Overview of Physical Scenarios}
\label{sec:scenarios}
We validate the 2D Fokker--Planck solver and analyze the problem against stochastic sLLG simulations across three physically distinct regimes of symmetry breaking:

\begin{itemize}
    \item \textbf{Scenario A, STT with minimal broken symmetry (PMA + $\varepsilon'$):} A spin-transfer torque device with perpendicular magnetic anisotropy (PMA) where azimuthal symmetry is broken only by the spin-polarization asymmetry parameter $\varepsilon' = 0.003$ \cite{donahue_object-oriented_2015}. This case demonstrates the transition from quasi-axisymmetric ($\varepsilon' \to 0$) to weakly asymmetric behavior.

    \item \textbf{Scenario B, STT with moderate broken symmetry (in-plane assist field $h_x = 0.12$):} The same STT device driven by an explicit in-plane external field, creating full azimuthal asymmetry and exposure of 2D transport pathways.

    \item \textbf{Scenario C, SOT with intrinsic broken symmetry (field-like torque with $\mathbf{h}_{\mathrm{norm}} = (0.15, 0.15, 0.15)$):} A spin-orbit torque device where azimuthal symmetry is broken inherently by the geometry of the field-like torque component, even without explicit in-plane assist fields. The competing damping-like and field-like torques, combined with three-dimensional external fields, create the most complex asymmetric landscape  \cite{sispad_modeling_2026}.
\end{itemize}

The three scenarios---parameterized in Table~\ref
{tab:parameters}---span the range from weak to strong 2D effects. By comparing FP predictions against $10^6$ stochastic trajectories, we isolate the discretization schemes' behavior across these regimes and demonstrate when 2D effects become physically essential.

\subsection{Analysis A: Péclet Number Regime Classification}

\begin{figure*}[!h]
\centering
\begin{subfigure}[b]{0.95\textwidth}
  \centering
  \includegraphics[width=1\textwidth]{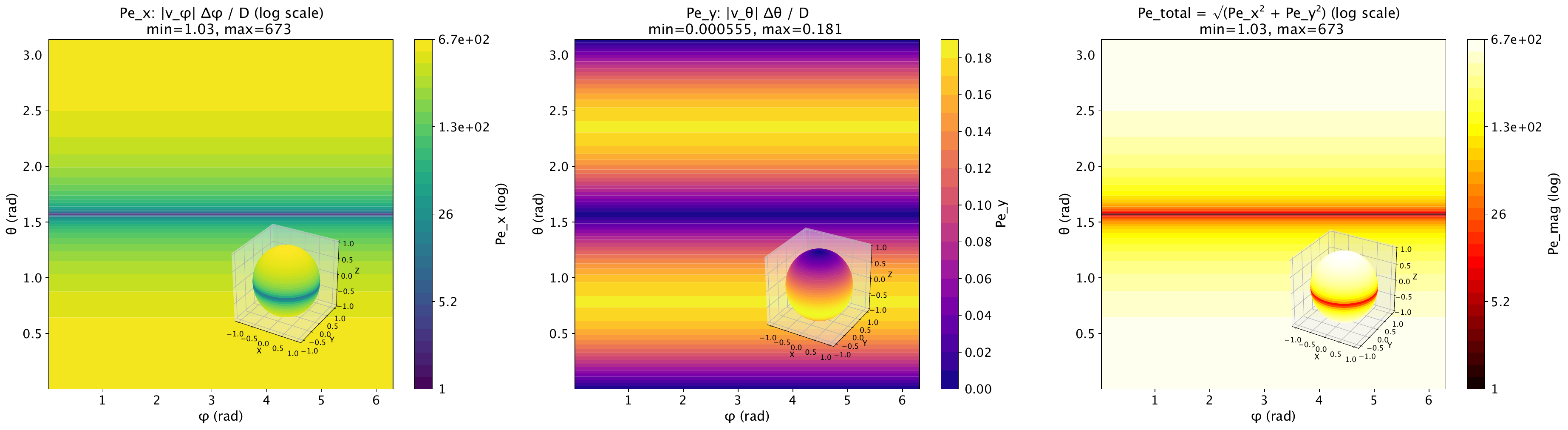}
  \caption{Péclet number distribution with $h_x = 0$ (quasi-axisymmetric case).}
  \label{fig:peclet_hx0}
\end{subfigure}
\begin{subfigure}[b]{0.95\textwidth}
  \centering
  \includegraphics[width=1\textwidth]{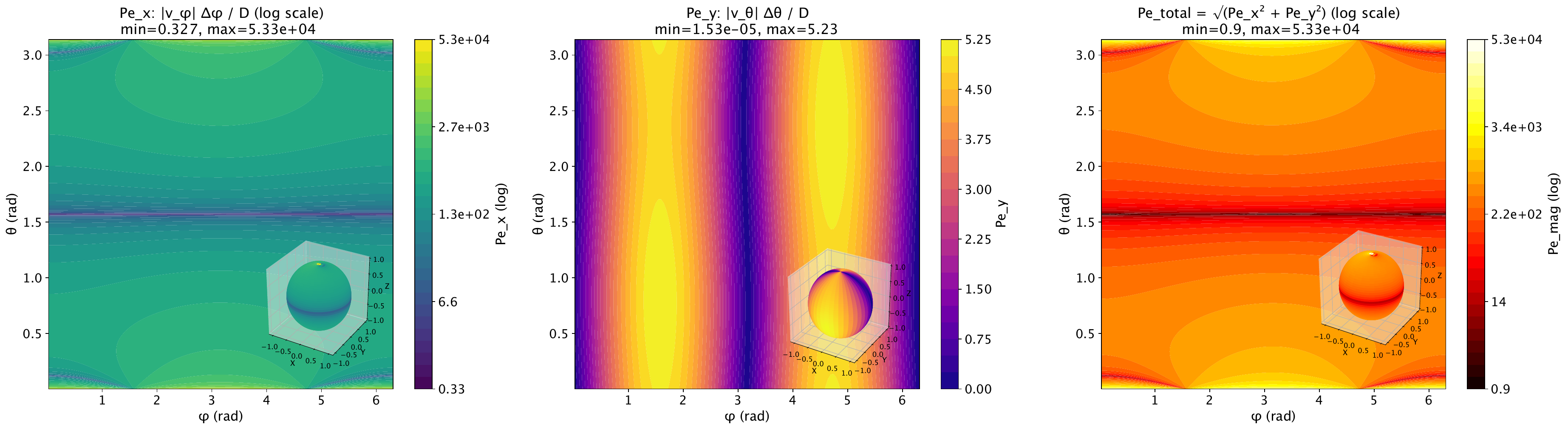}
  \caption{Péclet number distribution with $h_x = 0.12$ (broken symmetry case).}
  \label{fig:peclet_hx}
\end{subfigure}
\begin{subfigure}[b]{0.95\textwidth}
  \centering
  \includegraphics[width=1\textwidth]{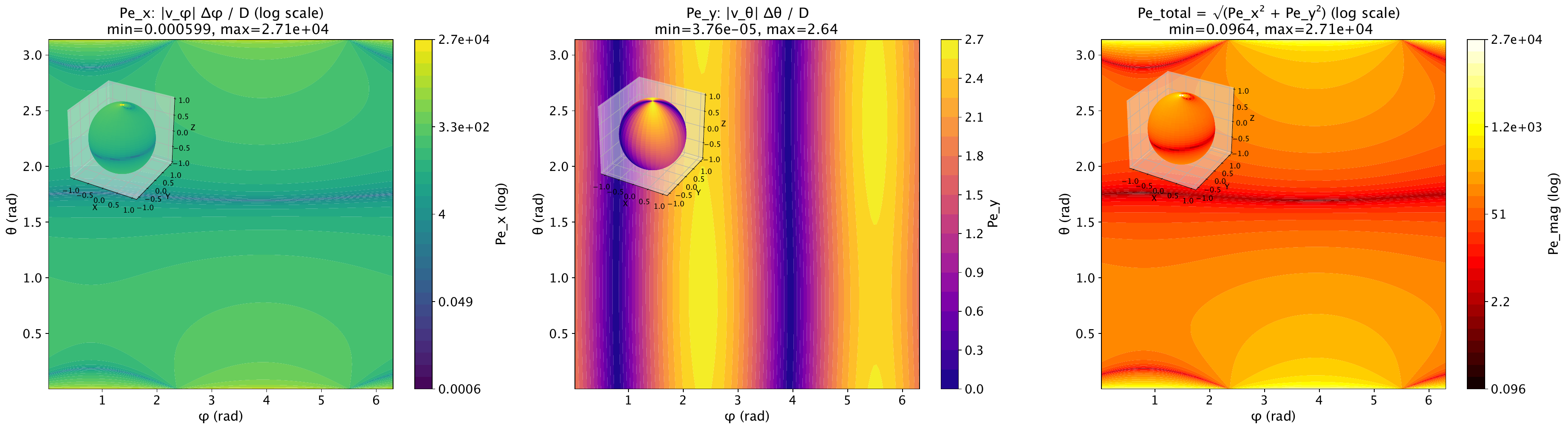}
  \caption{Péclet number distribution for SOT (field-like torque and external field $\mathbf{h}_{\mathrm{norm}} = [0.15, 0.15, 0.15]$), thus intrinsic broken symmetry.}
  \label{fig:peclet_sot}
\end{subfigure}
\caption{Spatial Péclet number distribution on the unit sphere: comparison across the three physical scenarios. Quasi-axisymmetric case (top) exhibits smoother, nearly uniform distribution. With broken azimuthal symmetry (middle and bottom), the full 2D transport structure emerges with localized high-Péclet regions.
}
\label{fig:peclet_comparison}
\end{figure*}

To characterize the transport regime across all three scenarios, we compute the local face Péclet number (Eq.~\ref{eq:peclet}) and analyze the spatial distribution and magnitude variations. Figure~\ref{fig:peclet_comparison} displays representative Péclet distributions for:

\begin{itemize}
    \item \textbf{Scenario A, STT with $h_x = 0$ (quasi-axisymmetric):} The Péclet field is nearly uniform along azimuthal lines, confirming that the solver almost reduces to an effective 1D problem, with the only discrepancies araising from $\varepsilon' \neq 0$. Ranges vary from $\mathcal{O}(1)-\mathcal{O}(10^3)$.

    \item \textbf{Scenario B, STT with $h_x = 0.12$ (moderate asymmetry):} Full 2D structure emerges. For representative parameters ($\alpha=0.0086$, $i=-2.35$), the azimuthal component dominates with $\mathrm{Pe}_\phi$ ranging from $10^{-1}$ to $\mathcal{O}(10^5)$, while $\mathrm{Pe}_\theta$ varies from $\mathcal{O}(10^{-5})-\mathcal{O}(10)$. High-Péclet regions are localized rather than global, occupying only a small fraction of the spherical domain.

    \item \textbf{Scenario C, SOT with competing torques:} The field-like torque creates broader azimuthal circulation, resulting in even larger spatial Péclet variations and stronger coupling between $(\theta, \phi)$ coordinates. $\mathrm{Pe}_\phi$ ranges from $\mathcal{O}(10^{-1})$ to $\mathcal{O}(10^5)$, and $\mathrm{Pe}_\theta$ from $\mathcal{O}(10^{-5})-\mathcal{O}(10)$
\end{itemize}

Consequently, excessive numerical stabilization in those localized regions can distort the net probability flow across the full domain and bias switching statistics. The key insight is that localized high-Péclet regions do not dominate the overall transport: most of the spherical domain remains diffusion-resolved, allowing classical flux-balance arguments to remain relevant.
To avoid large CFL values can cause solution smearing, we complement the Péclet analysis with advective CFL statistics.

\subsection{Analysis B: Mesh Resolution and Stability Requirements}

\begin{figure*}[!h]
\centering
\begin{subfigure}[t]{0.33\textwidth}
  \centering
  \includegraphics[width=\linewidth]{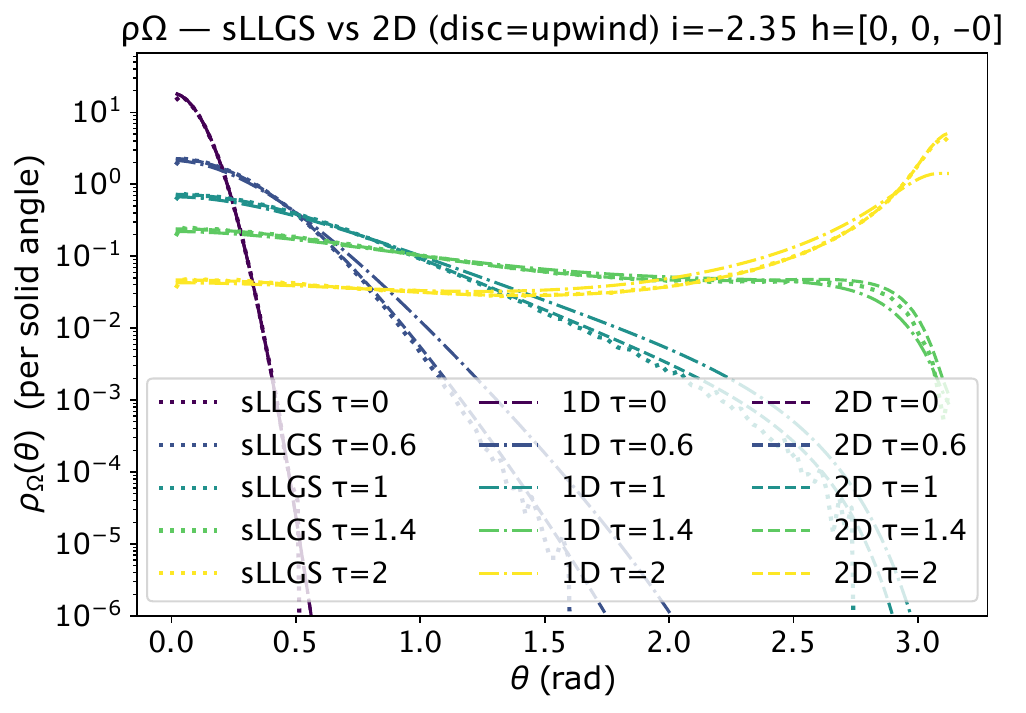}
  \caption{ STT $\rho(\tau)$, Upwind discretization.}
\end{subfigure}\hfill
\begin{subfigure}[t]{0.33\textwidth}
  \centering
  \includegraphics[width=\linewidth]{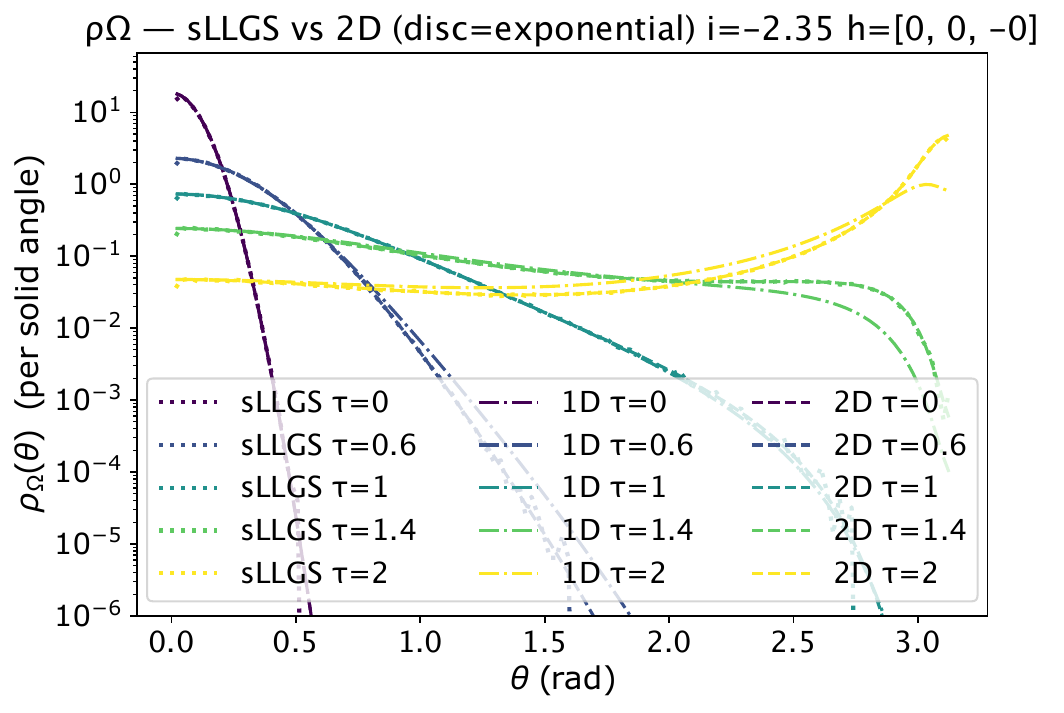}
  \caption{ STT $\rho(\tau)$, SG discretization.}
\end{subfigure}\hfill
\begin{subfigure}[t]{0.33\textwidth}
  \centering
  \includegraphics[width=\linewidth]{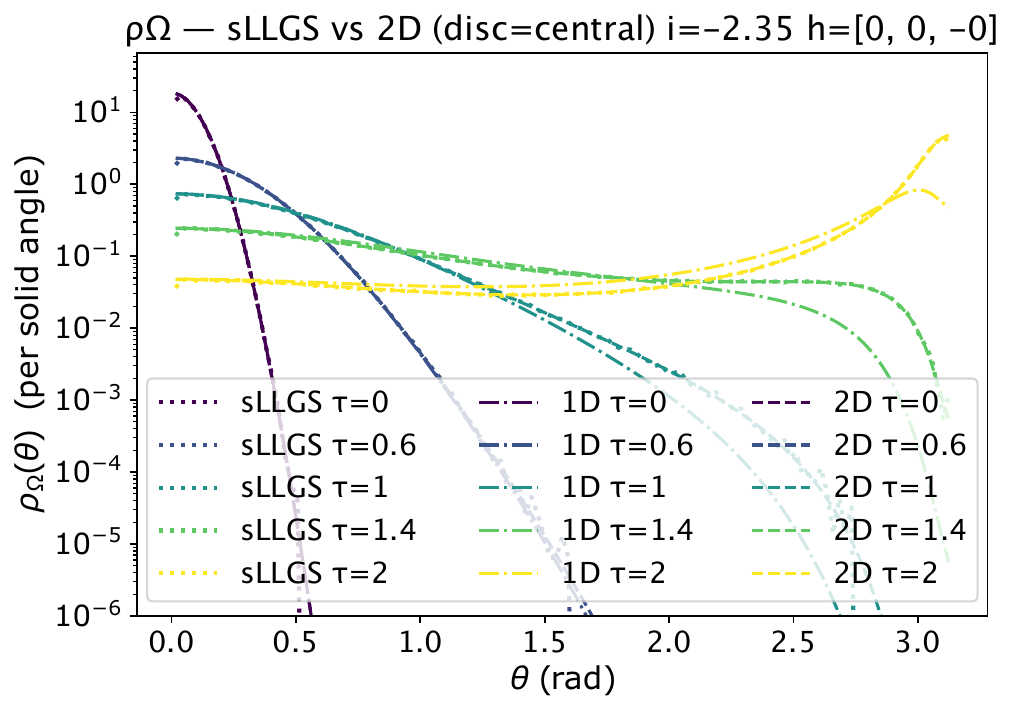}
  \caption{ STT $\rho(\tau)$, Central discretization.}
\end{subfigure}


\begin{subfigure}[t]{0.33\textwidth}
  \centering
  \includegraphics[width=\linewidth]{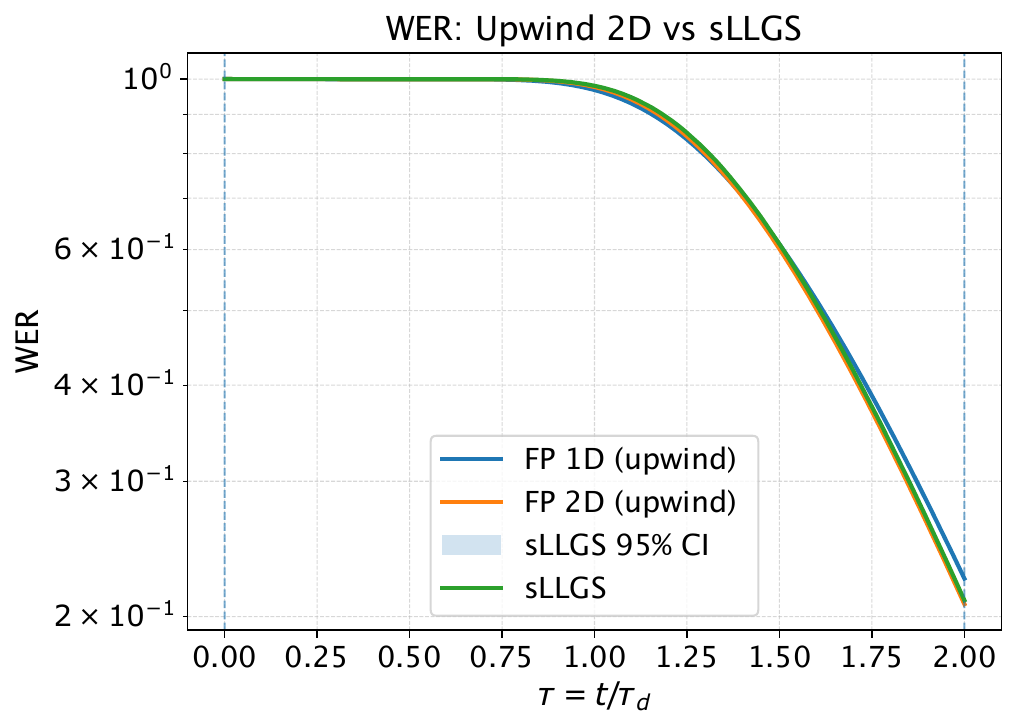}
  \caption{ STT WER, Upwind discretization.}
\end{subfigure}\hfill
\begin{subfigure}[t]{0.33\textwidth}
  \centering
  \includegraphics[width=\linewidth]{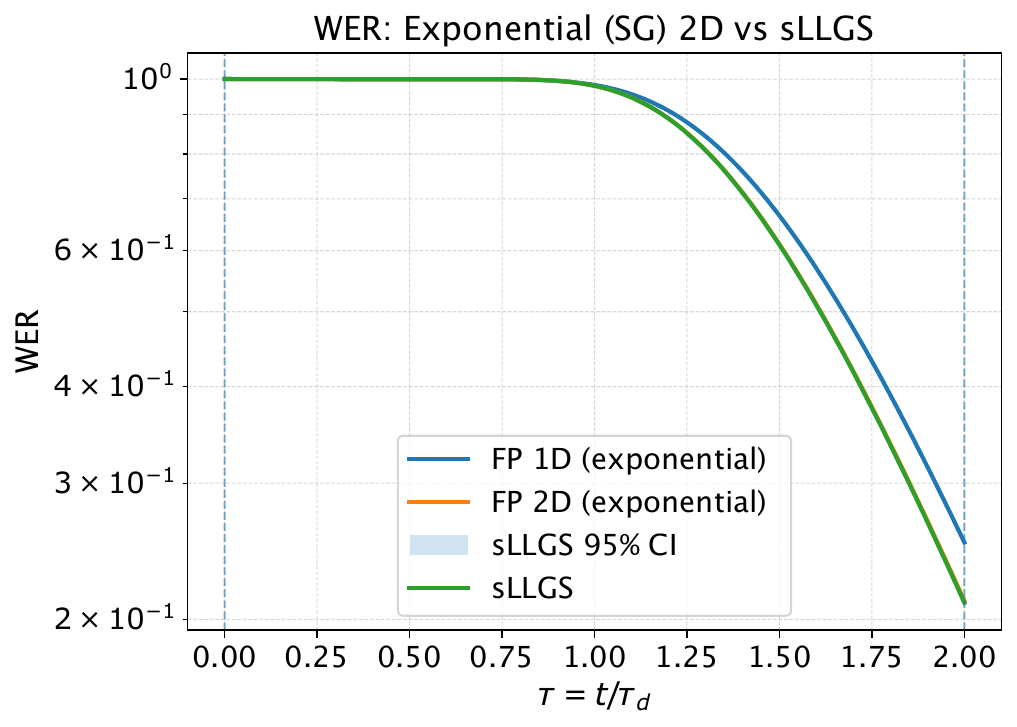}
  \caption{ STT WER, SG discretization.}
\end{subfigure}\hfill
\begin{subfigure}[t]{0.33\textwidth}
  \centering
  \includegraphics[width=\linewidth]{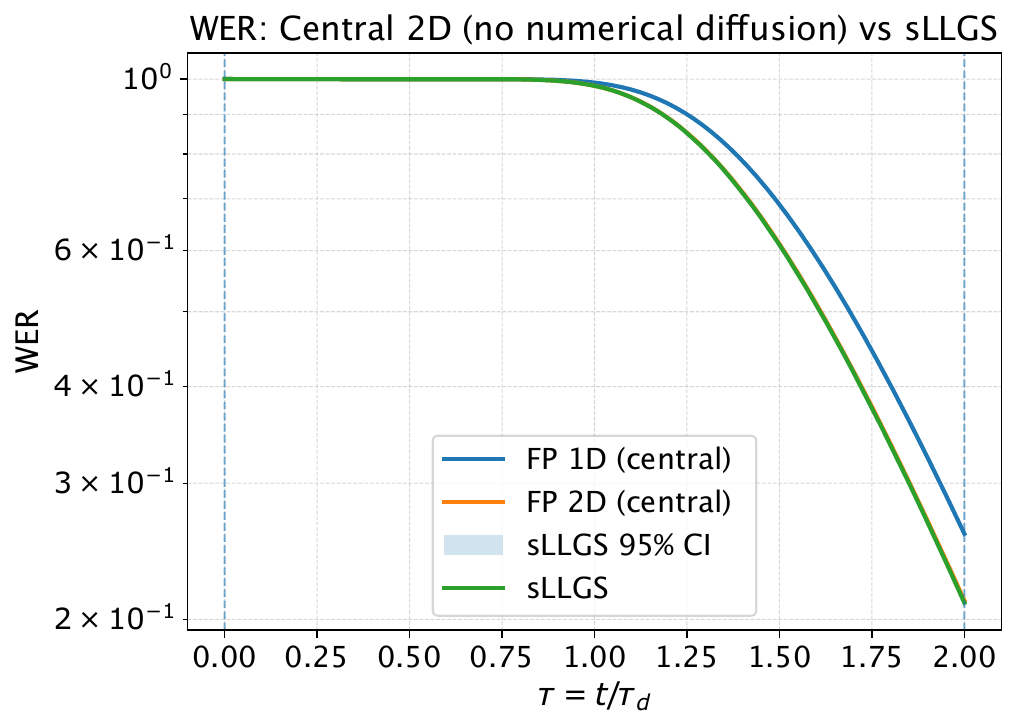}
  \caption{ STT WER, Central discretization.}
\end{subfigure}

\caption{STT device with in-plane assist field ($h_x = 0.0$).  Insets show probability density snapshots $\rho(\theta,\phi,\tau)$, and WER($\tau$) comparison between $10^6$-trajectory sLLG reference and 2D FP solutions using central, SG, and upwind fluxes.
The color of $\rho(\theta,\phi,\tau)$ distributions represents its normalized time evolution ($\tau=t/\tau_{d}$, $\tau_{d}=\frac{1+\alpha^2}{\alpha\gamma H_K^{\mathrm{eff}}}$) from its initial state (purple) to its final state (yellow). Given the quasi-axis-symmetry, every 2D discretization scheme is able to match the stochastic reference within Monte Carlo uncertainty. The presence of $\epsilon' \neq 0$ shows the 1D-solver inability to correctly track s-LLGS groundtruth.}
\label{fig:results_stt_hx_0}
\end{figure*}

\begin{figure*}[!h]
\centering
\begin{subfigure}[t]{0.33\textwidth}
  \centering
  \includegraphics[width=\linewidth]{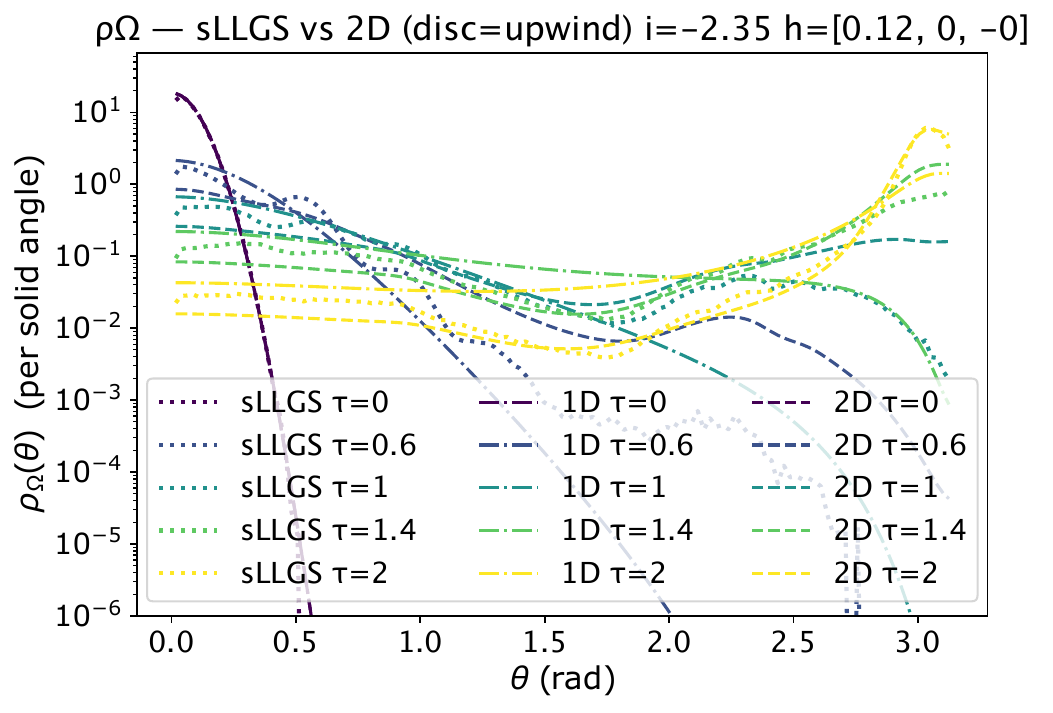}
  \caption{ STT $\rho(\tau)$, Upwind discretization.}
\end{subfigure}\hfill
\begin{subfigure}[t]{0.33\textwidth}
  \centering
  \includegraphics[width=\linewidth]{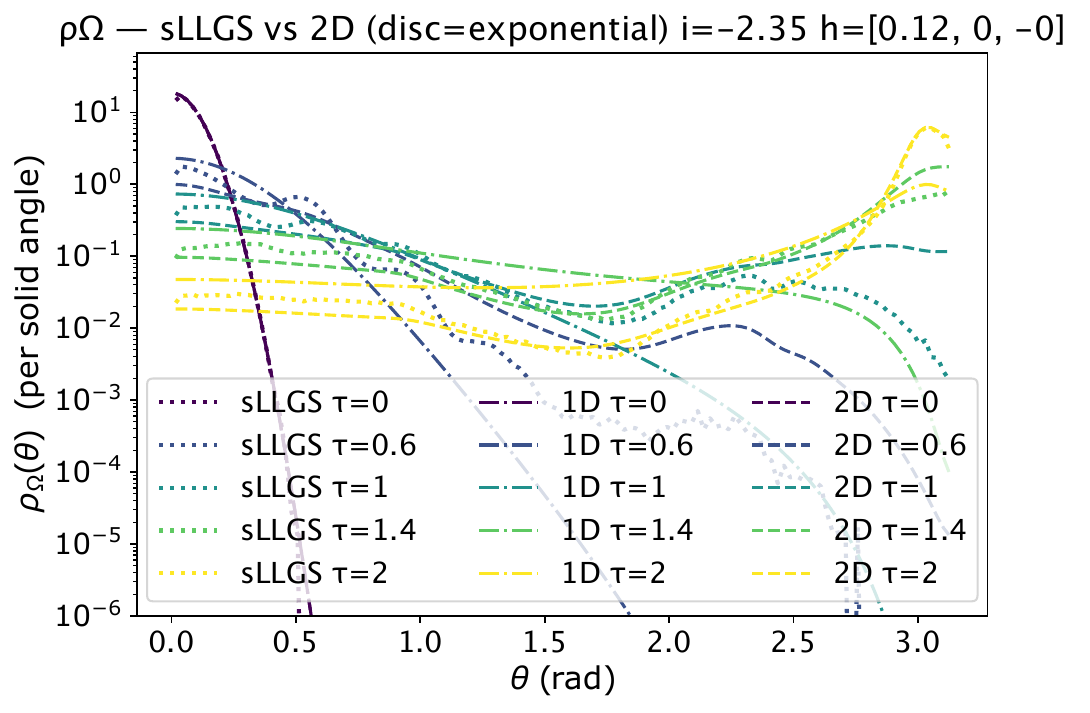}
  \caption{ STT $\rho(\tau)$, SG discretization.}
\end{subfigure}\hfill
\begin{subfigure}[t]{0.33\textwidth}
  \centering
  \includegraphics[width=\linewidth]{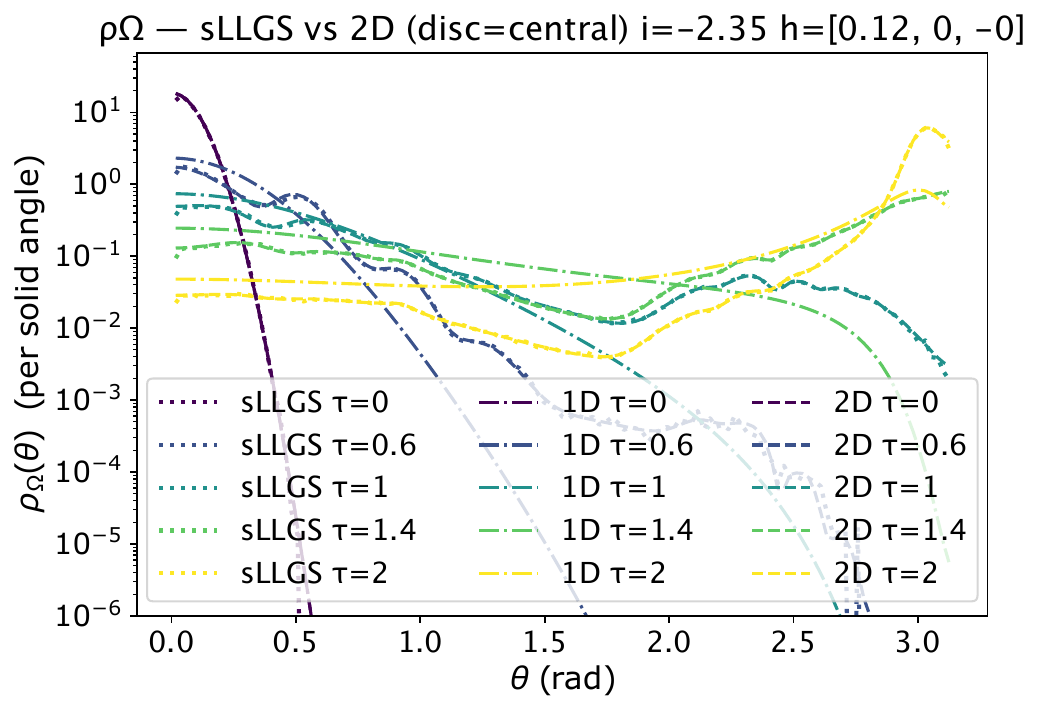}
  \caption{ STT $\rho(\tau)$, Central discretization.}
\end{subfigure}


\begin{subfigure}[t]{0.33\textwidth}
  \centering
  \includegraphics[width=\linewidth]{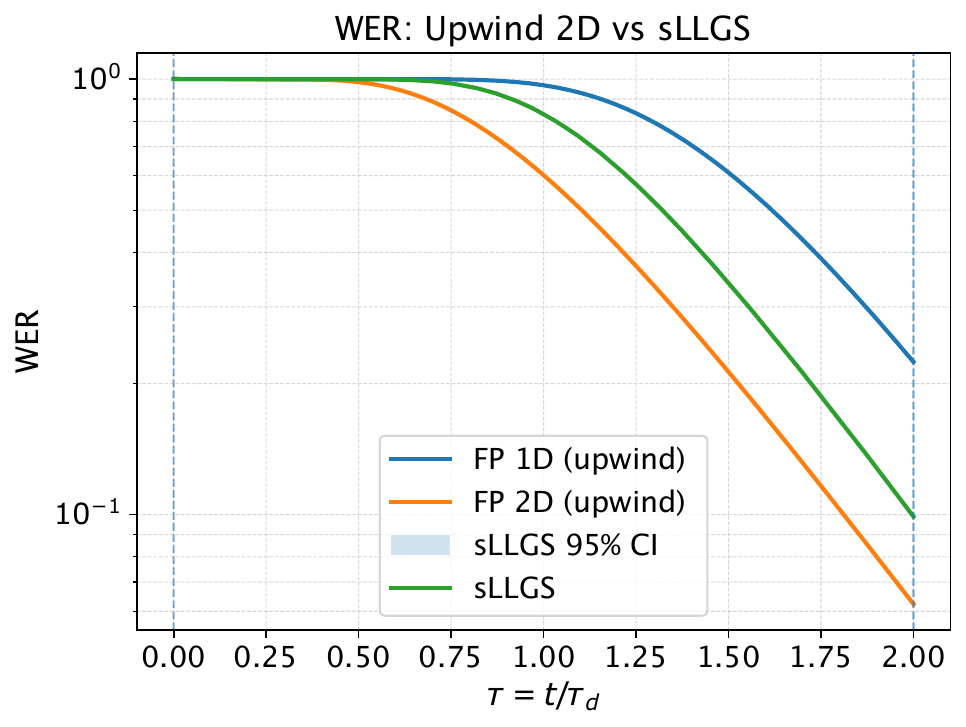}
  \caption{ STT WER, Upwind discretization.}
\end{subfigure}\hfill
\begin{subfigure}[t]{0.33\textwidth}
  \centering
  \includegraphics[width=\linewidth]{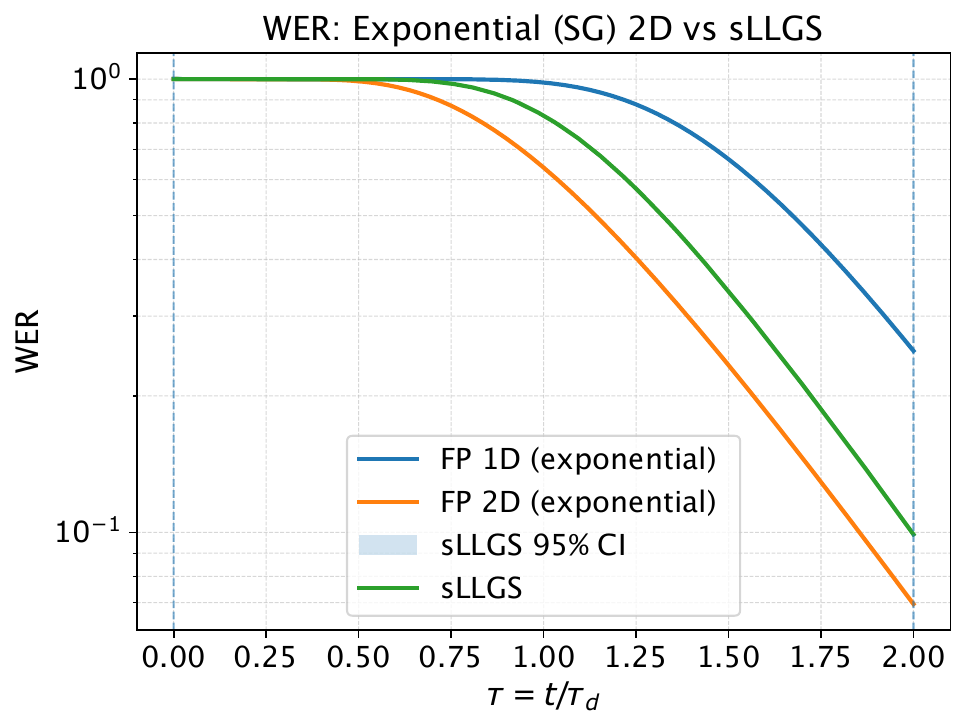}
  \caption{ STT WER, SG discretization.}
\end{subfigure}\hfill
\begin{subfigure}[t]{0.33\textwidth}
  \centering
  \includegraphics[width=\linewidth]{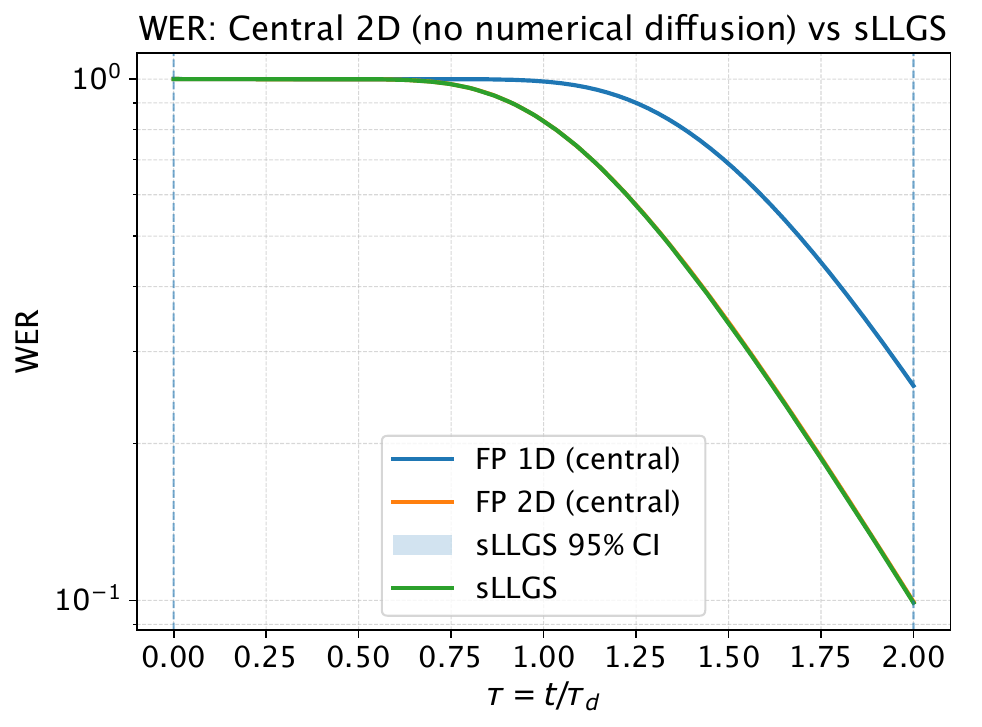}
  \caption{ STT WER, Central discretization.}
\end{subfigure}

\caption{STT device with in-plane assist field ($h_x = 0.12$). WER($\tau$) comparison between $10^6$-trajectory sLLG reference and 2D FP solutions using central, SG, and upwind fluxes. The presence of $\epsilon' \neq 0$, but more importantly $h_x \neq 0$, shows the 1D-solver inability to correctly track s-LLGS groundtruth. Central matches the stochastic reference within Monte Carlo uncertainty. SG and upwind predict earlier switching, reflecting altered phase and spreading of the probability pulse during barrier crossing. Insets show representative density snapshots $\rho(\theta,\phi,\tau)$, where central more accurately preserves the location and width of the switching probability packet.}
\label{fig:results_stt_hx_012}
\end{figure*}

\begin{figure*}[!h]
\centering
\begin{subfigure}[t]{0.33\textwidth}
  \centering
  \includegraphics[width=\linewidth]{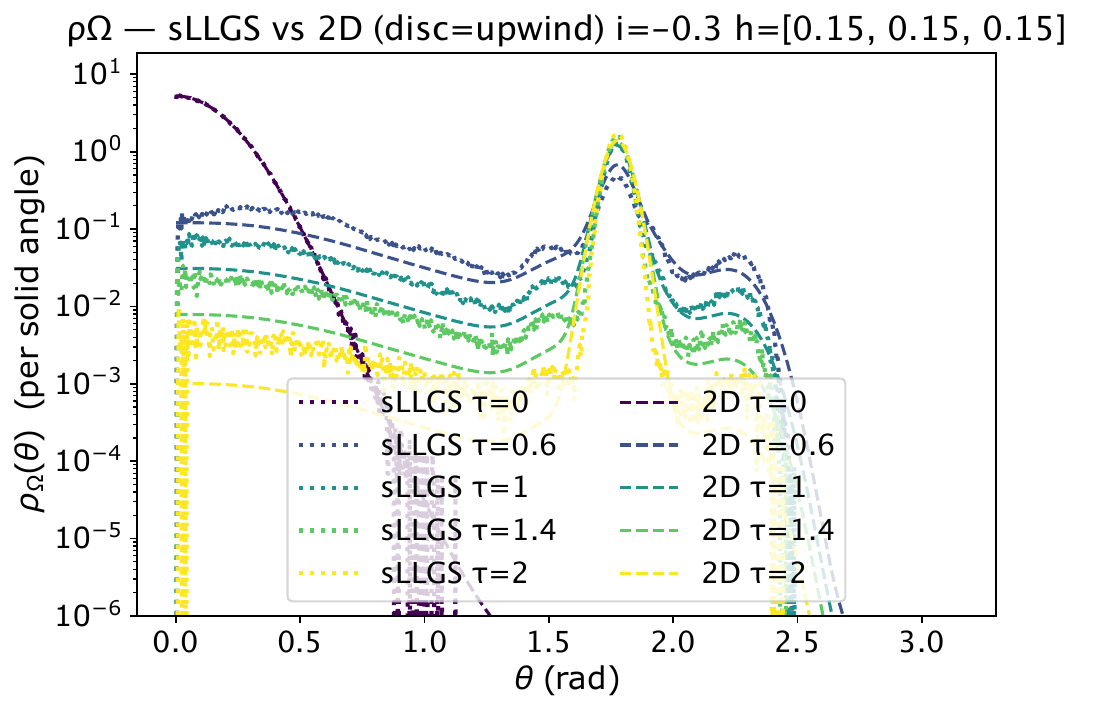}
  \caption{ SOT $\rho(\tau)$, Upwind discretization.}
\end{subfigure}\hfill
\begin{subfigure}[t]{0.33\textwidth}
  \centering
  \includegraphics[width=\linewidth]{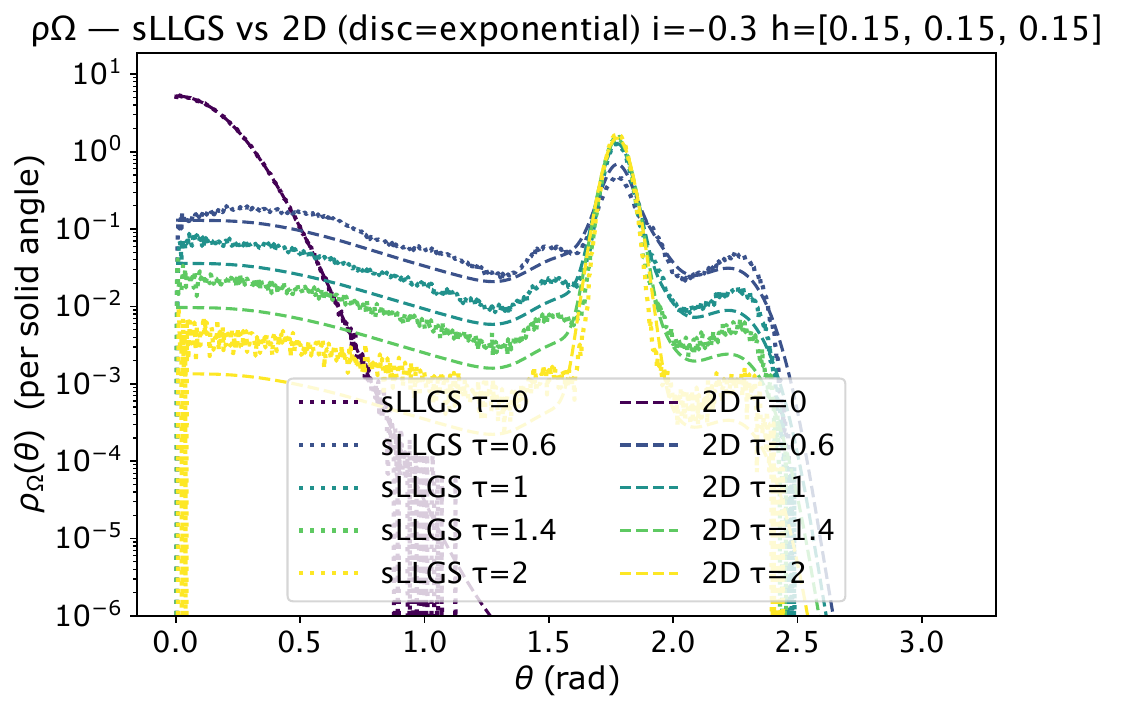}
  \caption{ SOT $\rho(\tau)$, SG discretization.}
\end{subfigure}\hfill
\begin{subfigure}[t]{0.33\textwidth}
  \centering
  \includegraphics[width=\linewidth]{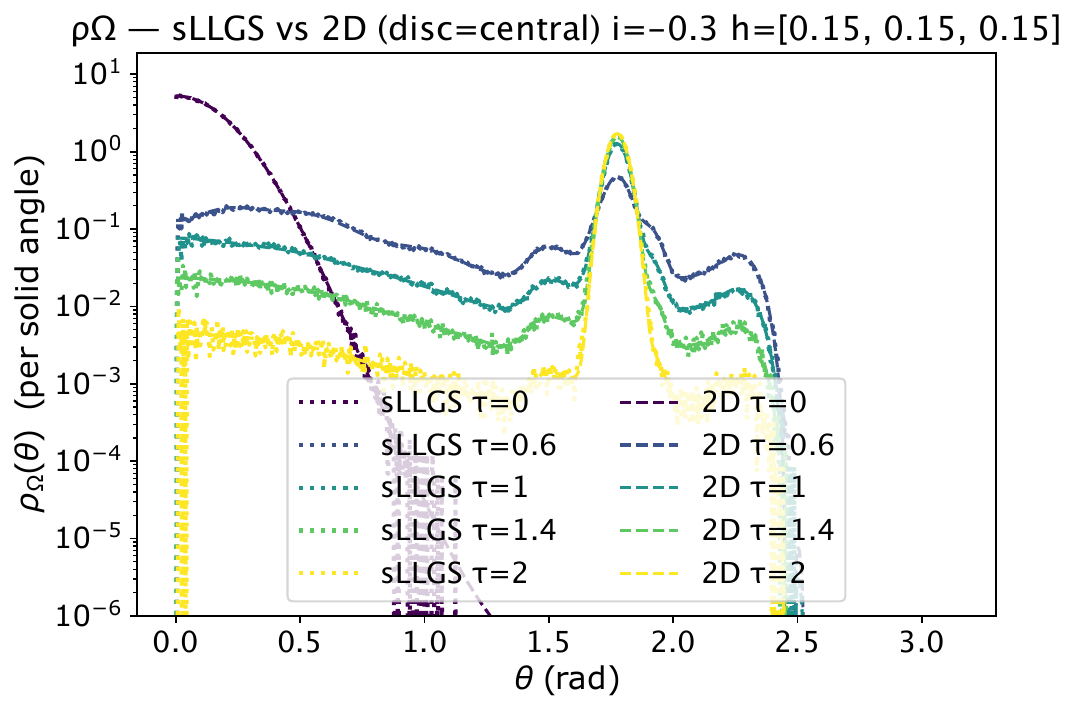}
  \caption{ SOT $\rho(\tau)$, Central discretization.}
\end{subfigure}


\begin{subfigure}[t]{0.33\textwidth}
  \centering
  \includegraphics[width=\linewidth]{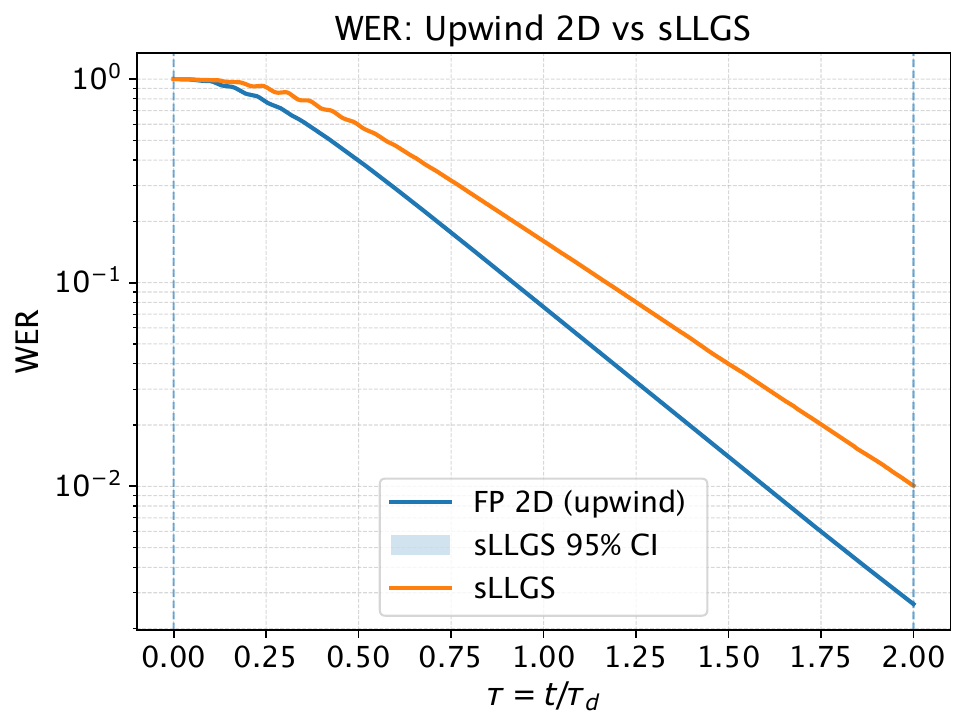}
  \caption{ SOT WER, Upwind discretization.}
\end{subfigure}\hfill
\begin{subfigure}[t]{0.33\textwidth}
  \centering
  \includegraphics[width=\linewidth]{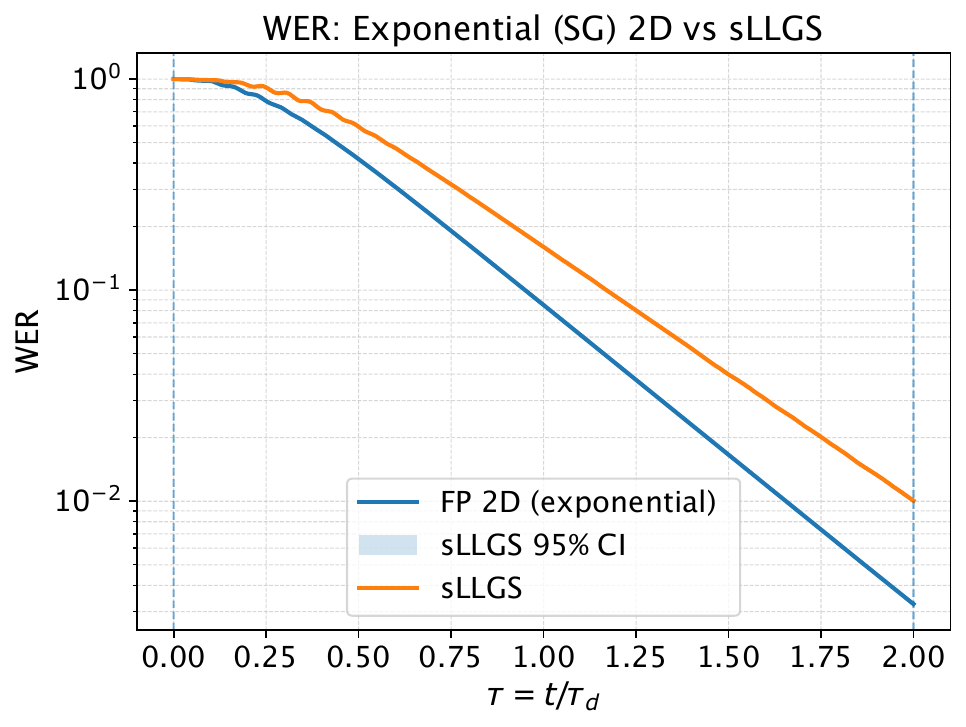}
  \caption{ SOT WER, SG discretization.}
\end{subfigure}\hfill
\begin{subfigure}[t]{0.33\textwidth}
  \centering
  \includegraphics[width=\linewidth]{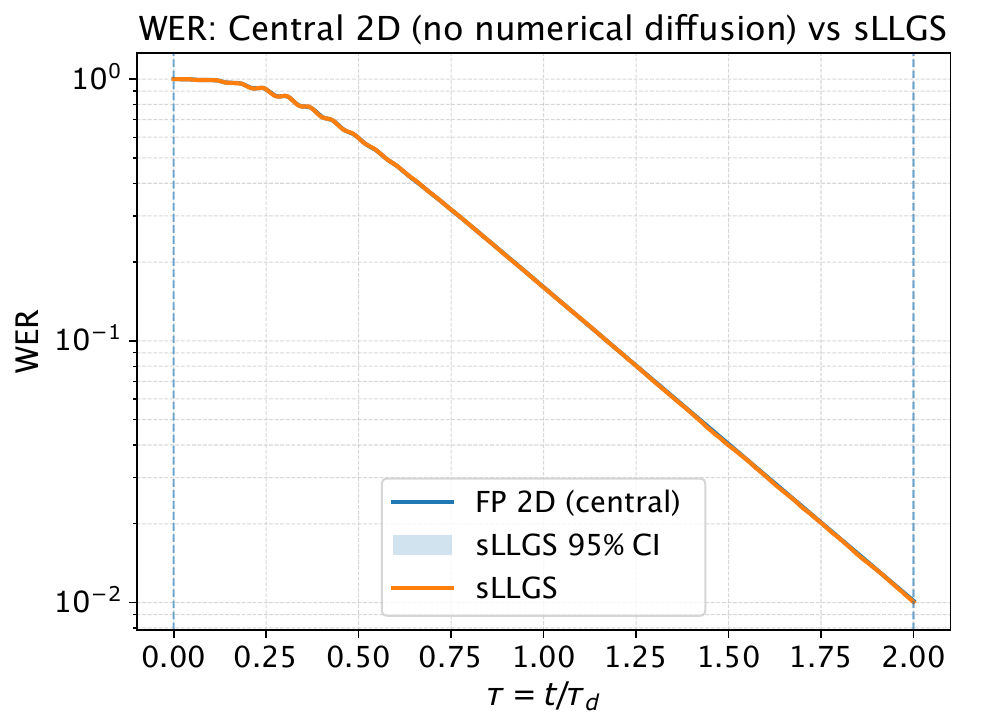}
  \caption{ SOT WER, Central discretization.}
\end{subfigure}

\caption{SOT device results. Top: stochastic switching trajectories showing the complex azimuthal spiraling induced by field-like torque. Bottom: WER($\tau$) comparison. Central FP overlaps the sLLG reference, whereas SG and upwind predict earlier switching with larger bias than the STT case. 1D reductions not represented as lacking strong azimuthal asymmetry capabilities.}
\label{fig:results_sot}
\end{figure*}

\begin{figure*}[!h]
\centering
\begin{subfigure}[t]{0.33\textwidth}
  \centering
  \includegraphics[width=\linewidth]{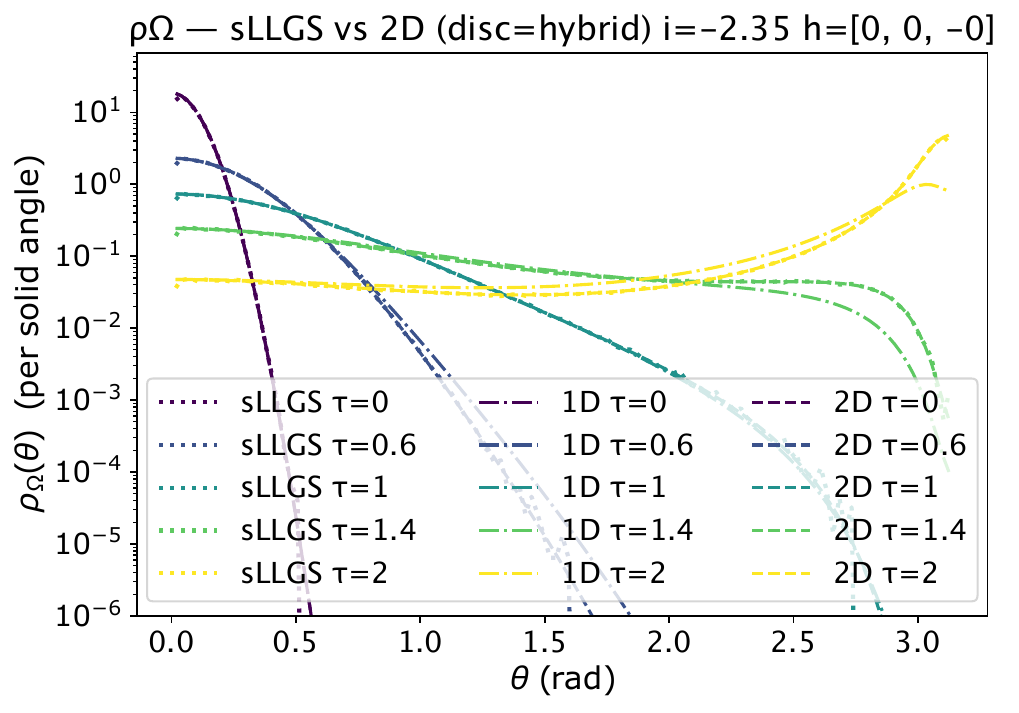}
  \caption{ STT $\rho(\tau)$, $\mathbf{h}_{\mathrm{norm}} = (0.0, 0.0, 0.0)$.}
\end{subfigure}\hfill
\begin{subfigure}[t]{0.33\textwidth}
  \centering
  \includegraphics[width=\linewidth]{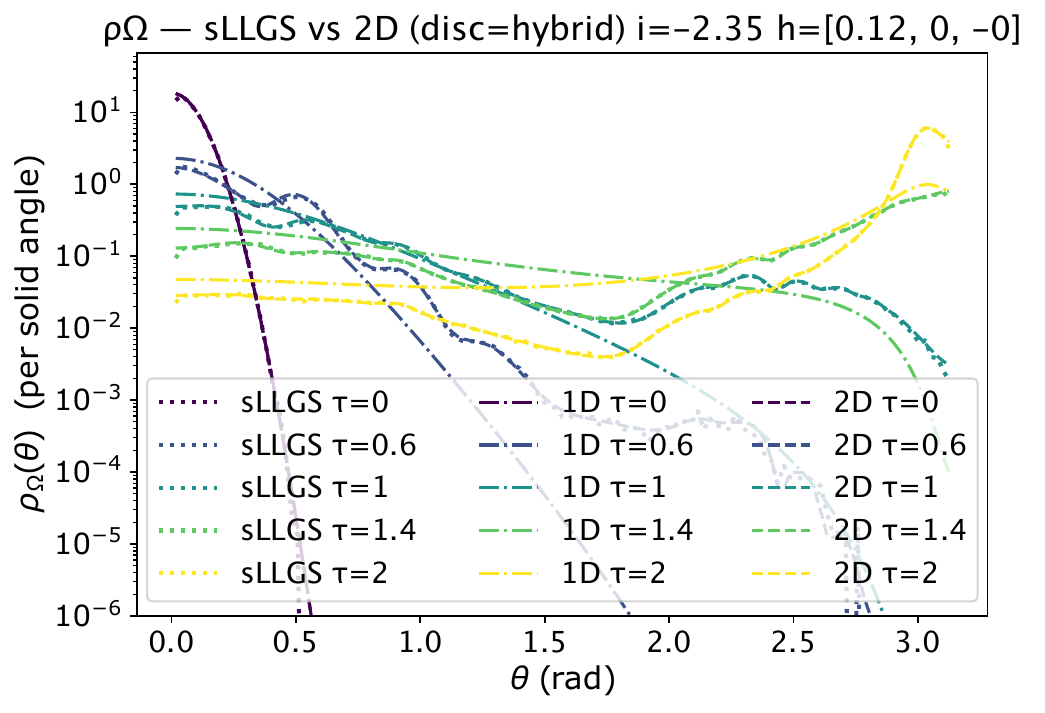}
  \caption{ STT $\rho(\tau)$, $\mathbf{h}_{\mathrm{norm}} = (0.12, 0.0, 0.0)$.}
\end{subfigure}\hfill
\begin{subfigure}[t]{0.33\textwidth}
  \centering
  \includegraphics[width=\linewidth]{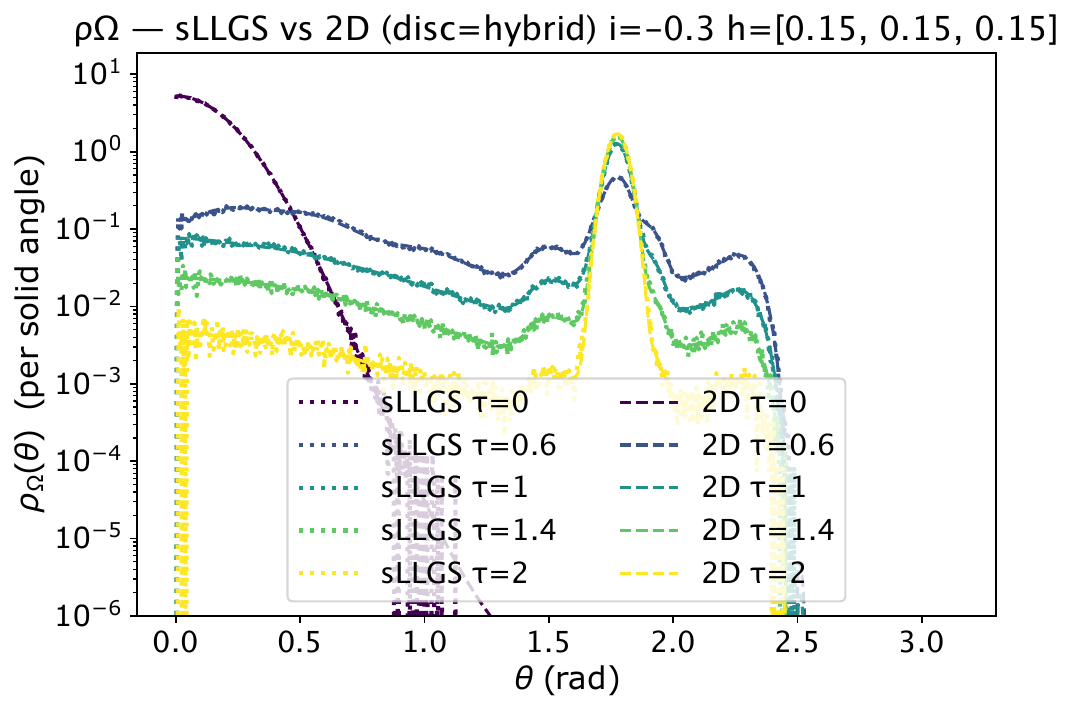}
  \caption{ SOT $\rho(\tau)$, $\mathbf{h}_{\mathrm{norm}} = (0.15, 0.15, 0.15)$.}
\end{subfigure}


\begin{subfigure}[t]{0.33\textwidth}
  \centering
  \includegraphics[width=\linewidth]{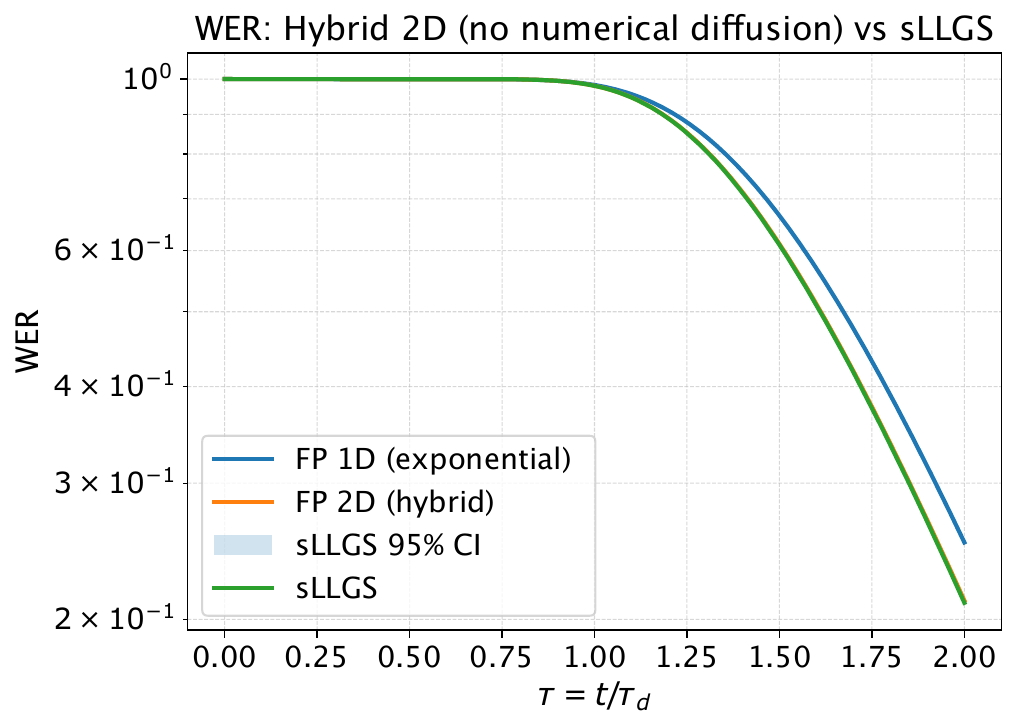}
  \caption{ STT WER, $\mathbf{h}_{\mathrm{norm}} = (0.0, 0.0, 0.0)$.}
\end{subfigure}\hfill
\begin{subfigure}[t]{0.33\textwidth}
  \centering
  \includegraphics[width=\linewidth]{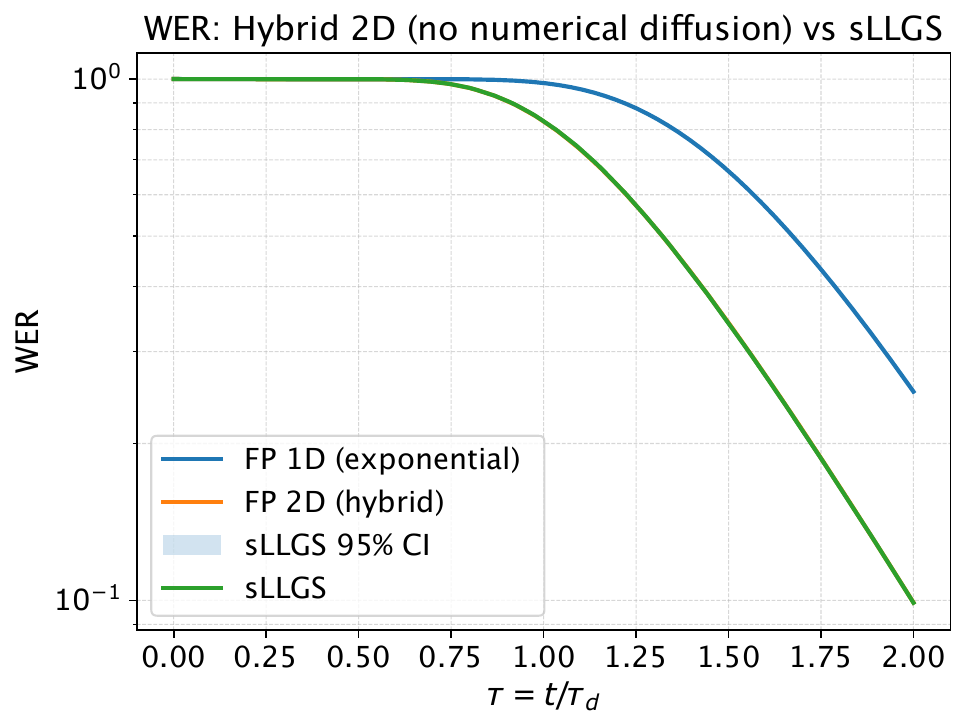}
  \caption{ STT WER, $\mathbf{h}_{\mathrm{norm}} = (0.12, 0.0, 0.0)$.}
\end{subfigure}\hfill
\begin{subfigure}[t]{0.33\textwidth}
  \centering
  \includegraphics[width=\linewidth]{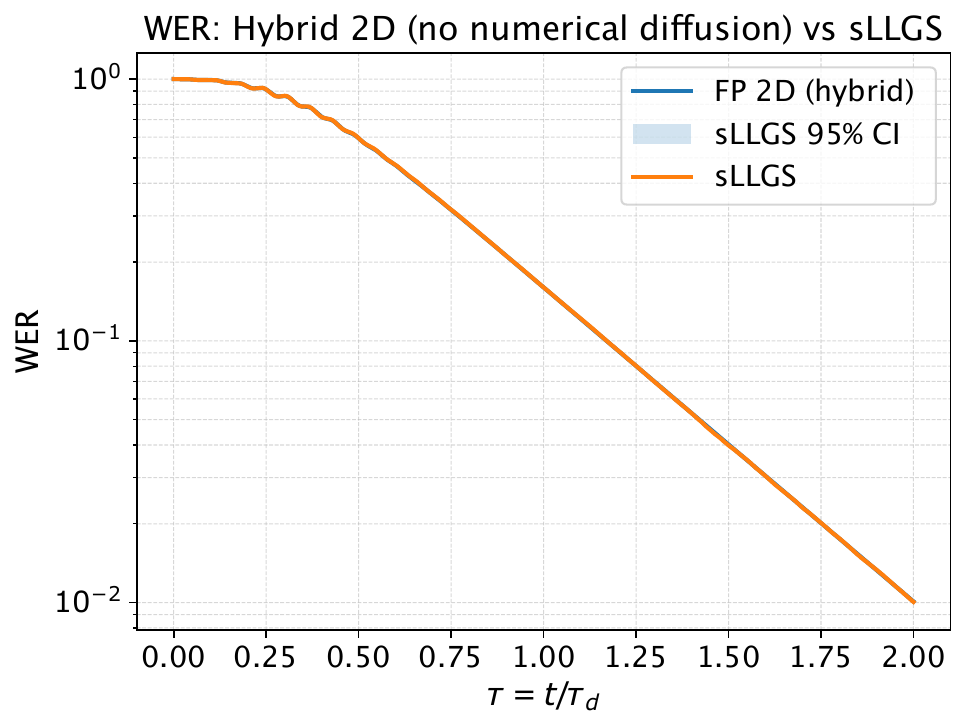}
  \caption{ SOT WER, $\mathbf{h}_{\mathrm{norm}} = (0.15, 0.15, 0.15)$.}
  \label{fig:results_hybrid}
\end{subfigure}

\caption{Hybrid discretization scheme shows consistent adaptive regime distribution, validating the blending strategy. These results demonstrate that the hybrid adaptive scheme provides a physics-preserving, automatically tuned discretization for magnetic devices with variable Péclet regimes.}
\label{fig:results_hybrid}
\end{figure*}

A key conclusion from the Péclet analysis is that mesh resolution must be adequate to resolve the localized high-drift regions without introducing artificial transport bias. For the spherical domain, we employ:

\begin{itemize}
    \item Angular resolution: $N_\phi \times N_\theta = 128 \times 1024$ finite-volume grid on ($\theta$, $\phi$), with uniform spacing spacing in $\theta$ and $\phi$.
    \item Temporal discretization: implicit Crank--Nicolson with $\Delta\tau = 10^{-3}$.
    \item CFL control. 
\end{itemize}

This resolution is sufficient to resolve the localized high-Péclet regions while maintaining diffusion-resolving coverage across most of the domain. Under these conditions, the reduced numerical dissipation of central differencing outweighs the monotonicity advantages of SG or upwind schemes, as we demonstrate below. When $\varepsilon' =0$ and $h_x = 0$ (quasi-axisymmetric case), the problem reduces to the conventional one-dimensional form, for which all three schemes produce nearly identical WER predictions after mesh refinement, confirming that discretization choice is irrelevant in the symmetric limit. Once $h_x \neq 0$, however, the full two-dimensional transport structure is exposed and differences between discretizations become measurable and physically important.

\subsection{Analysis C: Discretization Scheme Comparison}

\subsubsection{Scenario A, STT with Minimal Broken Symmetry (PMA + $\varepsilon'$)}

As a baseline, we first note that when $\varepsilon' \to 0$ and no external in-plane field is applied, the problem reduces to perfect axisymmetry. As seen in Figure~\ref{fig:results_stt_hx_0}, all three discretization schemes (central, Scharfetter--Gummel, upwind) produce nearly identical WER predictions, confirming that the physics is effectively 1D and scheme choice is irrelevant. With $\varepsilon' = 0.003$, the symmetry breaking is minimal, and WER differences between schemes remain small. This case validates that our solver correctly recovers classical 1D behavior in the symmetric limit.

\subsubsection{Scenario B, STT with Moderate Broken Symmetry (in-plane assist field $h_x = 0.12$)}

When an external in-plane field is applied, azimuthal symmetry is broken and the switching dynamics become intrinsically two-dimensional on the unit sphere. In this regime, and as seen in Figure~\ref{fig:results_stt_hx_012}, the numerical discretization of the Fokker--Planck (FP) flux operator becomes a critical determinant of write-error-rate (WER) accuracy. We validate against $10^6$-trajectory stochastic sLLG simulations.

The resulting WER($\tau$) predictions are:

\begin{itemize}
    \item \textbf{Central}: overlaps the sLLG reference within statistical uncertainty.
    \item \textbf{SG}: systematically shifted to shorter switching times.
    \item \textbf{Upwind}: exhibits the largest early-switching bias.
\end{itemize}

All simulations use identical drift physics, diffusion, boundary conditions, timestep, and mesh. The observed discrepancy therefore originates entirely from the spatial discretization of the transport operator. This key result demonstrates that central differencing more accurately captures the coupled $(\theta,\phi)$ transport pathways governing barrier crossing.

The discrepancy is not explained by classical one-dimensional Péclet-number arguments alone. Rather, it arises from the interplay of (i) rotational probability transport on the sphere, (ii) damping-driven drift, (iii) isotropic thermal diffusion, and (iv) metric terms associated with spherical coordinates. In the presence of broken symmetry, probability density evolves through coupled $(\theta,\phi)$ transport pathways, making the preservation of flux balance more important than monotonicity alone. In our tests, the central scheme most accurately preserves this balance, while SG and upwind fluxes introduce directional numerical bias that accelerates first-passage dynamics.

\subsubsection{Scenario C, SOT with Intrinsic Broken Symmetry (Field-Like Torque) and $\mathbf{H}_{\mathrm{ext}} \neq 0$}

We next consider a spin-orbit torque (SOT) device containing both damping-like ($\tau_{\rm DL}$) and field-like ($\tau_{\rm FL}$) torque components. Unlike STT, azimuthal symmetry is intrinsically broken by the torque geometry even without an explicit in-plane assist field. 
The three-dimensional external field $\mathbf{h}_{\mathrm{norm}} = (0.15, 0.15, 0.15)$ further enhances this asymmetry.
As reported in Figure~\ref{fig:results_sot}, the same trend observed in Scenario B is magnified here:

\begin{itemize}
    \item \textbf{Central}: agrees with the sLLG reference across the measured WER range.
    \item \textbf{SG / Upwind}: predict systematically faster switching, with larger discrepancy than the STT case.
\end{itemize}

The larger discrepancy arises because the field-like torque induces stronger azimuthal circulation and broader spreading in $\phi$, increasing sensitivity to transport discretization and elevating the importance of accurate 2D flux balance.

\subsection{Hybrid Adaptive Blending and Physical Consistency}
\label{sec:results_hybrid}
Having established that central differencing outperforms monotone schemes across all three scenarios, we now validate a hybrid adaptive discretization strategy (Section~\ref{sec:hybrid}) that automatically selects the optimal scheme based on local Péclet number. Figure~\ref{fig:results_hybrid} demonstrates that the hybrid adaptive blending scheme recovers the sLLG-validated WER predictions while providing diagnostic insight: in our geometries, for the scenarios in Section \ref{sec:scenarios} the scheme naturally adopts: 
\begin{itemize}
    \item Scenario 1: $\sim$86.9\% central differencing and $\sim$13.1\% blended region, with SG never selected.
    \item Scenario 2: $\sim$92.4\% central differencing and $\sim$7.6\% blended region, with SG never selected.
    \item Scenario 3: $\sim$785.5\% central differencing and $\sim$14.5\% blended region, with SG never selected.
\end{itemize}
This automatic validation confirms that central differencing is geometrically optimal for the coupled 2D transport governing spin-torque switching.

The superior performance of central differencing is also physically consistent with the structure of spin-torque switching dynamics. The dominant transport component is azimuthal and largely rotational on the unit sphere, particularly in the presence of in-plane assist fields or field-like torques. In this regime, preserving the correct phase and spreading of the probability packet during barrier crossing is more important than enforcing one-dimensional monotonicity at every face. Central differencing introduces the least artificial transport bias and therefore better preserves the coupled drift--diffusion balance governing first-passage switching statistics. By contrast, SG and upwind fluxes add directional numerical dissipation in localized high-Péclet regions, which can accelerate net probability leakage toward the switched state and lead to systematically optimistic (lower) WER predictions.

\subsection{Summary of Key Findings}

\begin{enumerate}
    \item \textbf{2D effects are essential for non-axisymmetric devices:} Reduced 1D models fail when symmetry is broken by in-plane fields (STT scenario 2) or field-like torques (SOT scenario 3). Full 2D transport resolution on the unit sphere is required for accurate WER predictions.

    \item \textbf{Discretization scheme choice matters:} Even with identical physics, mesh, and time stepping, different flux formulations produce measurably different switching statistics. Central differencing correctly preserves flux balance and reproduces stochastic reference WERs, while monotone schemes introduce systematic early-switching bias.

    \item \textbf{Localized high-Péclet regions do not govern overall accuracy:} The presence of localized high-Péclet zones does not justify global numerical stabilization. The majority of the spherical domain remains diffusion-resolved, making flux-balance preservation more important than monotonicity.

    \item \textbf{Hybrid adaptive schemes provide self-validated physics:} Automatic regime detection confirms that central differencing is geometrically optimal for the coupled $(\theta,\phi)$ transport in spin-torque devices, validating both the discretization choice and the problem structure itself.
\end{enumerate}

\section{Conclusions}

Fokker--Planck solvers are essential tools to study and guarantee the operation reliability in MRAM devices.
We have developed a 2D finite-volume Fokker--Planck solver and validated it against $10^6$-trajectory stochastic LLG for STT and SOT devices with broken symmetry~\cite{garcia2021compact,garcia2021fokker}. Key findings:

\begin{enumerate}
  \item \textbf{2D is necessary} for in-plane fields, field-like torques, SOT geometry, and asymmetric barriers. The 1D projection loses critical physics.
  \item \textbf{FVM guarantees mass conservation} to machine precision. Spherical metric factors are absorbed into mesh geometry, not pointwise divisions.
  \item \textbf{Discretization scheme matters}: Central gives unbiased diffusion ($D_{\rm eff} = D$) for Pe $< 2$; SG/upwind introduce systematic bias (+8\% to +50\% effective diffusion) that lowers predicted WER.
  \item \textbf{Customizable discretization} is essential: researchers can trade stability for accuracy, validate each choice, and adapt to their mesh resolution and Péclet regime. For magnetic simulations with variable Péclet regimes, the hybrid adaptive blending scheme is the recommended approach: it seamlessly balances accuracy across Pe ranges and naturally avoids schemes that over-diffuse, making it the preferred default for physics-preserving accuracy in magnetic memory simulations.
  \item \textbf{Every structure must be validated}: mesh resolution, time stepping, discretization scheme, boundary conditions. The solver's automated Pe reporting and scheme recommendation reduce error and increase confidence.
\end{enumerate}

This work establishes a foundation for accurate, unbiased WER prediction in next-generation STT/SOT-MRAM and provides tools for extending to more complex geometries and anisotropies.

The present core-framework is available for researchers at the repository \textcolor{red}{github.com/IMEC/PAPERUNDERREVIEW}.

\section*{Acknowledgments}
The authors would like to thank Dr. Maarten Van de Put, Dr. Ricardo Garcia Mayoral and Dr. Daniel Farrell for their helpful discussions and/or support during library development.

Work enabled by NanoIC pilot line (nanoic-project.eu), jointly funded by the Chips Joint Undertaking, through the EU’s Digital Europe (101183266) and Horizon Europe programs (101183277).

\bibliographystyle{IEEEtran}
\bibliography{IEEEcontrol, references}

@IEEEtranBSTCTL{IEEEexample:BSTcontrol,
  CTLname_url_prefix = "",
}

@inproceedings{garcia2021compact,
  author = {Garcia Redondo, Fernando and Prabhat, Pranay and Bhargava, Mudit and Dray, Cyrille},
  title = {A Compact Model for Scalable MTJ Simulation},
  booktitle = {IEEE Int. Conf. on Synthesis, Modeling, Analysis and Simulation Methods and Applications to Circuit Design},
  series = {SMACD},
  year = {2021},
  note = {Available: \url{https://arxiv.org/abs/2106.04976}}
}

@inproceedings{garcia2021fokker,
  author = {Garcia Redondo, Fernando and Prabhat, Pranay and Bhargava, Mudit},
  title = {A Fokker-Planck Solver to Model MTJ Stochasticity},
  booktitle = {European Solid-State Device Research Conf., ESSDERC},
  year = {2021},
  note = {\url{https://arxiv.org/abs/2106.12304}}
}

@book{strikwerda2004,
  author = {Strikwerda, John C.},
  title = {Finite Difference Schemes and Partial Differential Equations},
  edition = {2nd},
  publisher = {SIAM},
  year = {2004}
}

@book{morton2005,
  author = {Morton, K. W. and Mayers, D. F.},
  title = {Numerical Solution of Partial Differential Equations},
  edition = {2nd},
  publisher = {Cambridge University Press},
  year = {2005}
}

@article{scharfetter1969,
  author = {Scharfetter, D. L. and Gummel, H. K.},
  title = {Large-signal analysis of a silicon Read diode oscillator},
  journal = {IEEE Transactions on Electron Devices},
  volume = {16},
  number = {1},
  pages = {64--77},
  year = {1969},
  url={https://ieeexplore.ieee.org/document/1475609},
}

@book{selberherr1984,
  author = {Selberherr, Siegfried},
  title = {Analysis and Simulation of Semiconductor Devices},
  publisher = {Springer-Verlag},
  year = {1984},
  url = {https://link.springer.com/book/10.1007/978-3-7091-8752-4}
}

@book{gardiner2009,
  author = {Gardiner, C. W.},
  title = {Stochastic Methods: A Handbook for the Natural and Social Sciences},
  edition = {4th},
  publisher = {Springer},
  year = {2009}
}

@book{kloeden1992,
  author = {Kloeden, P. E. and Platen, E.},
  title = {Numerical Solution of Stochastic Differential Equations},
  publisher = {Springer-Verlag},
  year = {1992}
}

@article{kidger2021neuralsde,
  title={Neural {SDE}s as {I}nfinite-{D}imensional {GAN}s},
  author={Kidger, Patrick and Foster, James and Li, Xuechen and Oberhauser, Harald and Lyons, Terry},
  journal={International Conference on Machine Learning},
  year={2021},
  url           = {https://arxiv.org/abs/2102.03657}
}

@misc{donahue_object-oriented_2015,
    title = {{OOMMF} user's guide, version 2.1a},
    url = {https://math.nist.gov/oommf/software.html},
    doi = {10.18434/T4/1502495},
    language = {en},
    urldate = {2026-05-31},
    publisher = {National Institute of Standards and Technology},
    author = {Donahue, Michael},
    year = {2015},
}

@article{liu_high-accuracy_2025,
    title = {A {High}-{Accuracy} {STT}-{MTJ} {SPICE} {Model} {Based} on {Variable} {Parameters}},
    volume = {72},
    issn = {1557-9646},
    url = {https://ieeexplore.ieee.org/document/11003814/},
    doi = {10.1109/TED.2025.3566043},
    number = {7},
    urldate = {2026-05-31},
    journal = {IEEE Transactions on Electron Devices},
    author = {Liu, Haoyan and Gao, Shuchao and Chu, Chunshuang and Tian, Kangkai and Huang, Fuping and Zhang, Yonghui and Zhang, Zi-Hui},
    month = jul,
    year = {2025},
    pages = {3543--3549},
}

@article{Butler2012,
    title = {Switching {Distributions} for {Perpendicular} {Spin}-{Torque} {Devices} {Within} the {Macrospin} {Approximation}},
    volume = {48},
    issn = {0018-9464},
    url = {http://ieeexplore.ieee.org/document/6242414/},
    doi = {10.1109/TMAG.2012.2209122},
    number = {12},
    journal = {IEEE Transactions on Magnetics},
    publisher = {IEEE},
    author = {Butler, W. H. and Mewes, T. and Mewes, C. K. A. and Visscher, P. B. and Rippard, W. H. and Russek, S. E. and Heindl, R.},
    month = dec,
    year = {2012},
    pages = {4684--4700},
}

@article{Xie2017,
    title = {Fokker-{Planck} {Study} of {Parameter} {Dependence} on {Write} {Error} {Slope} in {Spin}-{Torque} {Switching}},
    volume = {64},
    issn = {00189383},
    doi = {10.1109/TED.2016.2632438},
    number = {1},
    journal = {IEEE Transactions on Electron Devices},
    publisher = {IEEE},
    author = {Xie, Yunkun and Behin-Aein, Behtash and Ghosh, Avik W.},
    year = {2017},
    pages = {319--324},
    url={https://ieeexplore.ieee.org/document/7797620},
}

@article{Torunbalci2018a,
    title = {Modular {Compact} {Modeling} of {MTJ} {Devices}},
    volume = {65},
    issn = {00189383},
    doi = {10.1109/TED.2018.2863538},
    number = {10},
    journal = {IEEE Transactions on Electron Devices},
    publisher = {IEEE},
    author = {Torunbalci, Mustafa Mert and Upadhyaya, Pramey and Bhave, Sunil A. and Camsari, Kerem Y.},
    year = {2018},
    pages = {4628--4634},
    url={https://ieeexplore.ieee.org/document/8443128},
}

@article{a_carrillo_finite-volume_2015,
    title = {A {Finite}-{Volume} {Method} for {Nonlinear} {Nonlocal} {Equations} with a {Gradient} {Flow} {Structure}},
    volume = {17},
    copyright = {https://www.cambridge.org/core/terms},
    issn = {1991-7120, 1815-2406},
    url = {https://www.global-sci.com/cicp/article/view/7533},
    doi = {10.4208/cicp.160214.010814a},
    urldate = {2026-06-01},
    journal = {Communications in Computational Physics},
    author = {A. Carrillo, José and Chertock, Alina and Huang, Yanghong},
    year = {2015},
}

@article{song_spin-orbit_2021,
    title = {Spin-orbit torques: {Materials}, mechanisms, performances, and potential applications},
    volume = {118},
    issn = {0079-6425},
    shorttitle = {Spin-orbit torques},
    url = {https://www.sciencedirect.com/science/article/pii/S0079642520301250},
    doi = {10.1016/j.pmatsci.2020.100761},
    urldate = {2026-06-03},
    journal = {Progress in Materials Science},
    author = {Song, Cheng and Zhang, Ruiqi and Liao, Liyang and Zhou, Yongjian and Zhou, Xiaofeng and Chen, Ruyi and You, Yunfeng and Chen, Xianzhe and Pan, Feng},
    month = may,
    year = {2021},
    pages = {100761},
}

@inproceedings{bhowmik_modeling_2025,
    title = {Modeling {SOT}-{Driven} {Domain} {Wall} {Motion} in {MTJ} {Switching}},
    url = {https://ieeexplore.ieee.org/document/11186379/},
    doi = {10.1109/SISPAD66650.2025.11186379},
    urldate = {2026-06-03},
    booktitle = {{Int.} {Conf.} on {Sim.} of {Sem.} {Proc.} and {Dev.} {SISPAD}},
    author = {Bhowmik, Trisha and Xiang, Y. and Gama Monteiro, M. and Rao, Siddharth and García-Redondo, F. and Van Houdt, J. and Temst, K.},
    year = {2025},
}

@inproceedings{sispad_modeling_2026,
    title = {Analytical SOT MRAM Modeling: From Deterministic to
Fokker-Planck based Stochastic Switching},
    booktitle = {{Int.} {Conf.} on {Sim.} of {Sem.} {Proc.} and {Dev.} {SISPAD}},
    author = {Bhowmik, Trisha and Gama Monteiro, M. and Rao, Siddharth and  Xiang, Y. and Van Houdt, J. and Temst, K. and Sankar Kar, Gouri and García-Redondo, F.},
    year = {2026},
}

@inproceedings{meeren_magnetic_2024,
    title = {Magnetic {Immunity} of {STT}-{MRAM}: {External} {Magnetic} {Field} {Orientation} {Impact} on {Writing} {Reliability}},
    issn = {2156-017X},
    shorttitle = {Magnetic {Immunity} of {STT}-{MRAM}},
    url = {https://ieeexplore.ieee.org/document/10873572/},
    doi = {10.1109/IEDM50854.2024.10873572},
    urldate = {2026-06-03},
    booktitle = {2024 {IEEE} {Int.} {Electron} {Devices} {Meeting} ({IEDM})},
    author = {Meeren, N. Vander and Van Beek, S. and Monteiro, M.G. and Garcia-Redondo, F. and Chatterjee, J. and Kumar, A. and Wostyn, K. and Couet, S. and Verbauwhede, I.},
    month = dec,
    year = {2024},
    note = {ISSN: 2156-017X},
    pages = {1--4},
}

@article{jiang_hybrid_2016,
    title = {Hybrid central-upwind finite volume schemes for solving the {Euler} and {Navier}–{Stokes} equations},
    volume = {72},
    issn = {08981221},
    url = {https://linkinghub.elsevier.com/retrieve/pii/S089812211630476X},
    doi = {10.1016/j.camwa.2016.08.022},
    language = {en},
    number = {9},
    urldate = {2026-06-03},
    journal = {Computers \& Mathematics with Applications},
    author = {Jiang, Zhen-Hua and Yan, Chao and Yu, Jian and Li, Yansu},
    month = nov,
    year = {2016},
    pages = {2241--2258},
}

\end{document}